\documentclass[prd, 12pt]{revtex4}
\usepackage{dcolumn}

\usepackage[pdftex]{color,graphicx,hyperref}
\usepackage{amsmath}
\usepackage{xfrac}
\usepackage{amssymb}
\usepackage{graphicx} 
\usepackage{booktabs} 
\usepackage{pdfsync}
\usepackage{verbatim}
\usepackage{latexsym}
\usepackage{dsfont}
\usepackage{epstopdf}
\usepackage{bm}

\def\sec#1{\section{#1} }
\def\ssec#1{\subsection{#1} }
\def\sssec#1{\subsubsection{#1} }
\def\R{\\\\} 
\def\({\left(}
\def\){\right)}
\def\[{\left[}
\def\]{\right]}
\def\a{\alpha}

\def\f#1#2{\frac{#1}{#2}}
\def\g{\gamma}
\def\d{\partial}
\def\de{\delta}
\def\De{\Delta}

\def\del{\nabla}

\def\e{\eta}

\def\k{\kappa}
\def\l{\lambda}
\def\L{\Lambda}
\def\m{\mu}
\def\n{\nu}
\def\o{\omega}

\def\p{\pi}
\def\r{\rho}
\def\s{\sigma}
\def\t{\tau}
\def\th{\theta}

\def\<{\langle}
\def\>{\rangle}

\newcommand{\ket}[1]{\left| #1 \right>} 
\newcommand{\bra}[1]{\left< #1 \right|} 

\providecommand{\abs}[1]{\lvert#1\rvert}

\definecolor{orange}{rgb}{1,0.5,0}
\definecolor{test}{rgb}{.5,0.5,.5}

\begin{document}

\title{Brane Stabilization and \emph{Regionality} of Extra Dimensions}

\author{David M. Jacobs\footnote{Email address: david.m.jacobs@case.edu}, Glenn D. Starkman\footnote{Email address: glenn.starkman@case.edu}, and Andrew J. Tolley\footnote{Email address: andrew.j.tolley@case.edu}}%

\affiliation{Center for Education and Research in Cosmology and Astrophysics}
\affiliation{Department of Physics and Institute for the Science of Origins, \\Case Western Reserve University}


\begin{abstract}

Extra dimensions are a common feature of  beyond the Standard Model physics.  
In a braneworld scenario, 
  local physics on the brane can depend strongly on the brane's location within the bulk. 
  Generically, the relevant properties of the bulk manifold for the physics on/of the brane are neither local nor global, 
  but depend on the structure of finite regions of the bulk,  even for locally homogeneous and isotropic bulk geometries. 
  In a recent work we considered various mechanisms (in a braneworld context) to stabilize the location of a brane within bulk spaces of non-trivial topology.  In this work we elaborate on and generalize that work by considering additional bulk and brane dimensionalities as well as different boundary conditions on the bulk scalar field that provides a Casimir force on the brane, providing further insight on this effect.\\  

In $D=2+1$ ($D=5+1$) we consider both local and global contributions to the effective potential of a 1-brane (4-brane) wrapped around both the 2-dimensional hyperbolic horn and  Euclidean cone, which are used as toy models of an extra-dimensional manifold.  We calculate the total energy due to brane tension and elastic energy (extrinsic curvature) as well as that due to the Casimir energy of a bulk scalar satisfying both Dirchlet and Neumann boundary conditions on the brane.  In some cases stable minima of the potential are found that result from the competition of at least two of the contributions.  Generically, any one of these effects may be sufficient when the bulk space has less symmetry than the manifolds considered here.  We highlight the importance of the Casimir effect for the purpose of brane stabilization.

\end{abstract}

\maketitle

\tableofcontents
 
\sec{Introduction}

Several outstanding problems remain in fundamental physics. To name a few:  unification of the gauge interactions of the Standard Model with the gravitational interaction, the enormous hierarchy between the observed gravitational and weak scales, $M_\text{pl}/M_\text{weak}\approx 10^{17}$, as well as the hierarchy among fundamental fermion masses (e.g. $m_\t/m_e \approx 3500$), and  the smallness of an apparent cosmological constant, $\L$, where $\sqrt{\L} \approx H_0 \approx 10^{-33}$eV$ \approx 10^{-44}M_\text{weak}$\textsuperscript{\footnotemark[1]}\footnotetext[1]{As usual we have set $\hbar=c=1$.}.

Extra spatial dimensions might exist and may help to explain some, if not all of these issues:  they are an essential feature of string theory;  ``large" extra dimensions have been employed in attempts to explain the weak (e.g. \cite{ArkaniHamed:1998rs,Antoniadis:1998ig,Randall:1999ee}) and flavor  (e.g. \cite{ArkaniHamed:1999dc}) hierarchies; there have also been attempts to address the dark energy problem via infra-red modifications of gravity that employ extra dimensions (e.g. \cite{Dvali:2000hr, Dvali:2007kt,deRham:2007xp}).

Common experience and laboratory experiments tell us we don't appear to live in more than three spatial dimensions; if more exist they should be hidden in some way.  One explanation might be that the Standard Model fields are confined to submanifolds of lower dimension or perhaps the extra dimensions are compactified (i.e. the manifold has non-trivial topology) to length scales that have remained inaccessible by experiments.  The actual situation may be a combination of both. Furthermore, since topology is a \emph{global} property of the space-time manifold it cannot be determined by the \emph{local} Einstein's equations -- interestingly, though, topology can have a large influence on local physical processes within the manifold.

In what follows below we will consider scenarios where the full space-time manifold is the direct product of 4D Minkowksi space-time, ${\cal M}^4$,  (or similarly, Friedmann-Robertson-Walker) with a d-dimensional extra  manifold, $\Omega^d$, but certainly the discussion could extend to other scenarios.  With $d=1$ only the topology needs to be specified, as the geometry is trivial. From experience we know geometry can play an important role as well, therefore we begin our discussion with $d=2$.


Let us first consider a cylinder with topology ($\mathbb{R} \times S^1$). It can be obtained from Euclidean 2-space (${E}^2$) by modding out by a group ($\Gamma$) of 1-dimensional discrete translations. The local geometry remains the same as the covering space (i.e. homogeneous and isotropic), however the presence of fields indicates that some (global) symmetry has been broken.  For example, the components of field momenta are quantized in the compact dimension, while they remain continuous in the infinite dimension. That is, local experiments are sensitive to the global structure of the manifold. Even so, \emph{no special places exist on this space}, as the results of such experiments are insensitive to where on the manifold they are performed.

\begin{figure}
  \begin{center}
    \includegraphics[scale=.8]{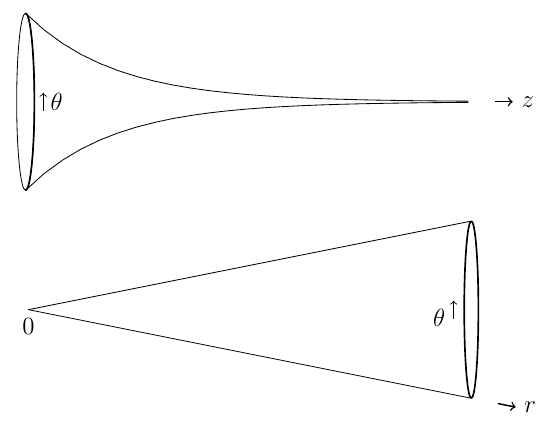}
  \end{center}
\caption{Horn (top) and Cone (bottom) topologies.}
\label{Horn_and_Cone}
\end{figure}

This should be contrasted with the two-dimemsional hyperbolic horn (${\cal H}^{2}/\Gamma$), obtained from the hyperbolic 2-space ${\cal H}^{2}$ by modding out by the same $\Gamma$ (see Figure \ref{Horn_and_Cone}). Fields on this space indicate that both rotational \emph{and} translational symmetry are broken;  field modes with non-zero excitation in the compact dimension are highly inhomogeneous in $z$ and are doubly exponentially suppressed beyond the point on the manifold where the ``circumference" of the compact dimension becomes smaller than the mode wavelength. In other words, modes of a certain momentum can only be sent a finite distance down the horn. Since translation invariance is broken, there exists a notion of absolute position and the results of local experiments will vary along the length of the horn. The main lesson here is that even if the local geometry of a manifold is homogeneous and isotropic, the physics on it will generally not be.

The physically relevant scale in this example is the horn circumference, a quantity that is sensitive to non-local characteristics of the manifold, yet is insensitive to the whether or not the manifold is actually entirely a horn that extends to infinity with infinite volume. For this reason we term such a quantity \emph{regional}, a quantity that could not have been calculated simply using local geometric information, yet did not require complete knowledge of the manifold's structure. The hyperbolic horn is of great utility because it has a lot of the salient features of more generic manifolds, in particular the ``cuspy" regions of compact hyperbolic manifolds (CHMs)\textsuperscript{\footnotemark[2]}\footnotetext[2]{CHMs provide an appealing geometric solution for the hierarchy problem (see \cite{Kaloper:2000jb}); they could be considered a hyperbolic, $d>1$, version of models discussed in \cite{ArkaniHamed:1998rs} or \cite{Randall:1999ee}, wherein all but one modulus is fixed. CHM's have also found use in string theory (see e.g. \cite{Orlando:2010kx}).}. It therefore seems that what may appear as a global effect can actually be attributed to the regional structure of the manifold one considers.  

 Another simple but non-trivial topology is that of a cone, a two-dimensional Euclidean space with a ``wedge" removed from it. It is obtained by identifying $\th \leftrightarrow \th + 2\p(1-\de)$, when working in polar coordinates, $r$ and $\th$ (see Figure \ref{Horn_and_Cone}). Topologically, the space is given by $\mathbb{R}^2-\{0\}$, i.e. the point at the tip of the cone is removed from the manifold. Again, translation invariance is broken, as the distance from the tip of the cone is imprinted on the modes.  The cone offers a complementary example to the horn as it has zero curvature and is a better approximation to regions of manifolds that are geometrically flat and end in a vertex.


We would like to focus our attention on braneworld scenarios and the local physical consequences of regional properties of the bulk manifold. They would affect, e.g. vacuum energy of a brane itself, as it is generically proportional to the inverse of the brane volume, which can vary as the brane is moved through the manifold. Furthermore, the amplitude for brane fields to interact with fields in the bulk (e.g. the gravitational field) would also be sensitive to the position of the brane, as the structure of bulk modes are generally inhomogeneous. Finally, as we shall explore in this work, there are local forces on branes arising from e.g. the Casimir effect of bulk fields interacting with the brane. {\bf }

Before exploring any interesting phenomenological prospects, however, we must ask what determines the brane's location within the bulk and what ensures it is stabilized.   From an effective 4D perspective the brane's position within the bulk appears as a series of massless scalar fields, one for each of the brane's codimenensions. Observational constraints on fifth forces tell us that such fields should have a mass $\gtrsim 10^{-3}$ eV \cite{Kapner:2006si} if they couple to matter with gravitational strength.  In any case, stability is a reasonable requirement of any system and thus the brane's effective potential should have a stable minimum in which it sits. Stabilizing a brane in this context should be contrasted with other mechanisms (e.g.  \cite{Goldberger:1999uk}) that have been used in the context of stabilizing the moduli of the bulk -- here we are concerned with determining and stabilizing the brane position \emph{within} a bulk that is assumed to be stable.

In this work we focus on different local (geometric) and global 
contributions to the brane potential and find that they depend on the manifold's non-trivial structure in different ways. From classical intuition, we expect that a brane's non-zero tension will provide an energy proportional to the brane volume.  Furthermore, there will generally be an elastic energy associated with the brane's extrinsic curvature and how it couples to the intrinsic bulk curvature.

Due to local interactions with the brane, inhomogeneities in bulk field modes will be induced -- this will be approximated here by an effective boundary condition on the bulk field(s) at the location of the brane.  Therefore if global translation invariance of the bulk manifold is broken, the vacuum energy of bulk fields (a global quantity) will depend on the location of the brane, providing a location-dependent force, i.e. the Casimir effect \cite{Casimir:1948dh}. Likewise, the Casimir energy of Standard Model fields living on the brane can play a role if the brane's topology is non-trivial.  Calculation of Casimir energies is subtle because it strongly depends on the bulk and boundary geometry, topology, dimensionality, field type, and boundary conditions.  The magnitude of the effect can be estimated on dimensional grounds;  however in order to get the overall sign and precise magnitude of the energy, a full calculation usually must be performed.  Having explicit analytic expressions for the fields modes (as one has on the horn and cone) makes this task much more tractable.

Though using Casimir energies to stabilize bulk moduli is not a new idea (see e.g \cite{ Garriga:2000jb, Ponton:2001hq, Greene:2007xu, Elizalde:2002dd}), to our knowledge they have not been used to stabilize the location of a single brane within the bulk nor have the regional properties of manifolds been exploited for their use. We would especially like to emphasize the role of the Casimir effect because, in the case of a 3-brane embedded in a higher-dimensional bulk, the brane geometry is trivial, as are the corresponding energies. The Casimir effect may then be the only mechanism available to stabilize the brane's location in this case.  

\begin{figure}
\begin{center}
    \includegraphics[scale=1]{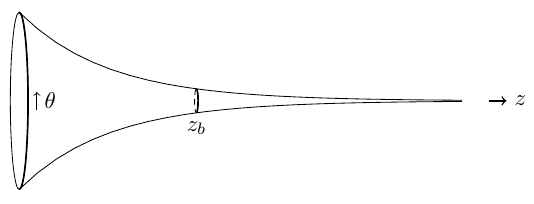}
\end{center}
\caption{A partial embedding diagram of the horn. The codimension-1 brane is pictured with coordinate $z_b$. The manifold is infinite in extent in the $\pm z$ directions.} 
\label{fig_horn}
\end{figure}

In summary, the gradients of the total energy result in a net \emph{force}, which can stabilize the location of a brane.  In this work, we generalize the results of \cite{Jacobs:2012ph}, confining our attention to a codimension-1 brane wrapped around either the 2-dimensional hyperbolic horn or the Euclidean cone, but extending our analysis to additionally include the Neumann boundary condition, as well as analogous calculations in $D=2+1$ (i.e without the Minkowski dimensions).  On both geometries (in $D=5+1$) we suppose that all Standard Model fields are confined to a codimension-1 brane as pictured in Figure \ref{fig_horn} or \ref{fig_cone}, but that they are free to propagate in the compact (i.e. ``universal" \cite{Appelquist:2000nn}) dimension. This is possible if the ``circumference" of the brane is small enough so that the first excited ``Kaluza Klein" mode has not yet been made accessible by experiment.

We have calculated the total energy of these systems as the sum of tension and elastic energies (i.e. those local to the brane and associated with its geometry) as well as the (global) vacuum energy from a bulk scalar field, $\phi$, that satisfies a boundary condition on the brane. We also parametrize the vacuum energy of Standard Model fields confined on the brane. 
For each geometry, we first consider total space-time dimension $D=2+1$ then consider the addition of three Minkowski dimensions in $D=5+1$. While the 2+1 D cases are interesting for academic reasons, they also illuminate how the different contributions to the brane potential scale differently with the number of spatial dimensions as well as provide  examples of the how the explicit poles of the vacuum energy depend significantly on the even/oddness of $D$.



In section \ref{horn} we analyze the hyperbolic horn, considering both dimensionalities and boundary conditions.  There we discover, in both dimensionalities, that a competition between tension and bulk Casimir energies can result in a stable brane position, however only for the Dirichlet case.  In Section \ref{cone} we perform the analogous calculations for the cone. A stable minimum is possible for all boundary conditions in $D=2+1$ and can result from the competition between the tension and elastic energies (i.e. purely from geometric effects), but is also ensured by the Casimir energy from the scalar. In $D=5+1$ a global minimum occurs for the Dirichlet case while only a local minimum is possible for Neumann.  In Section \ref{conclusions} we conclude with a summary of results and discussion.


\sec{Hyperbolic Horn}\label{horn}

\ssec{Preliminaries}\label{section:prelims}
\sssec{The Model, Energy Contributions, and Bulk $\phi$ Solutions}

Here we will consider a general spacetime dimension $D=m+2+1$, where $m$ indicates the number of Minkowski spatial dimensions. That is, the 
the full spacetime manifold is  ${\cal M}^{m+1}\times {\cal H}^{2}/\Gamma$. The line element can be written in coordinates such that
\begin{equation*}
ds^2 = \e_{\m\n}^{(m+1)}dx^\m dx^\n  +e^{-2z/z_\star} z_\star^2 d\th^2 + dz^2 \,,
\end{equation*}
where we identify $(\th) \leftrightarrow (\th +2\p)$,  and we have chosen our coordiantes such that the horn circumference at $z=0$ is $2\p z_\star$. Furthermore, from here on we shall work in units where the intrinsic length scale of the horn, $z_\star \equiv 1$. The model is illustrated in Figure \ref{fig_horn}, suppressing any Minkowksi spatial dimensions, where the codimension-1 brane wraps around the compact direction and resides at coordinate $z_b$. In $D=5+1$ ($m=3$) we assume the Standard Model fields to be confined to the brane, free to propagate in the compact direction. We will now enumerate different contributions to the brane's effective potential on the horn.
\R\\
\emph{Energy Contributions}\\

There will be a contribution from a non-zero brane tension, $\s$, that provides an energy
\begin{equation}
E^{~}_{\text{ten}}=\int d^{1+m}x \sqrt{\abs{\g}} \s = 2\p e^{-z_b} V_{M}\s \,,
\end{equation}
where $\g_{\m\n}$ is the induced spatial metric on the brane, and the volume of the regulated Minkowski spatial slice is denoted by $V_{M}$. There will also be an ``elastic" energy contribution due to the extrinsic curvature of the brane, $K_{ab}$ (see Appendix \ref{heat_horn}). We write this as an expansion in scalars built from $K_{ab}$, namely
\begin{equation}
E^{~}_{\text{curv}}=\int d^{1+m}x \sqrt{\abs{\g}} \(h_1K^{2}+h_2K_{ab}K_{ab}+\dots\) = 2\p e^{-z_b} V_{M} \(h_1+h_2+\dots\)
\end{equation}
where the $h_i$ are parameters describing the energy cost for deforming the brane within the manifold, and $\dots$ represents other possible scalars\textsuperscript{\footnotemark[3]}\footnotetext[3]{In principle we should also include possible coupling between the extrinsic-  and Riemann curvature tensors, however in this geometry those terms appear with the same $z_b$-dependence, thus at this level they are indistinguishable from the terms with only extrinsic curvature.} that are higher order in $K_{ab}$.  Note that the effect of $\s$ and the $h_i$ are completely indistinguishable, at least as far as the total energy of the system is concerned. Without loss of generality, then, we shall set $h_i=0$ and encode all geometric effects in $\s$. It is already clear that these effects behave monotonically with $z_b$, and are thus insufficient to stabilize the brane by themselves, but as we will see, the Casimir energies from the brane or bulk can provide a complementary opposing force. 

The Casimir (or vacuum) energy in the bulk scalar $\phi$ on a constant-$t$ hypersurface is
\begin{equation}
E_0=\int \sqrt{-g}~ d \Sigma \bra{0} T_{00}^{(\phi)}\ket{0} 
\end{equation}
from which one obtains the canonical result $E_0=\f{1}{2}\sum_{\bf i} \o_{\bf i}$, where ${\bf i}$ is a general mode index. It is well known that this sum is infinite;  to extract the physically relevant energy we employ the zeta-function regularization technique (see e.g. \cite{Bordag:2009zz}) and write


\begin{equation}\label{zeta_fn_reg_E_0}
E_0(s)=\f{\m^{2s}}{2}\sum_{\bf i} \o_{\bf i}^{1-2s}
\end{equation}  
where $\m$ is a renormalization scale.  The generalized sum is finite for large enough $s$, which is analytically continued to zero.  In some systems the terms in the energy that determine the Casimir \emph{force} remain finite even in the limit $s\to 0$, while in others they diverge and must be dealt with explicitly; in either case it is ultimately the renormalization of geometric parameters in the full theory (including gravity) that remedies the situation.

In the $D=5+1$ model there will also be a vacuum energy associated with the Standard Model fields living on the brane, sensitive to the size of the brane in the compact dimension.  Though we will not calculate this effect explicitly, by dimensional considerations we expect its associated 4D energy density to scale approximately inversely proportional to the fourth power of this length scale, namely 
\begin{equation}
\r_{0, SM} \simeq \f{\k_{SM}}{\(2\p e^{-z_b}\)^4}
\end{equation}
where $\k_{SM}$ is a dimensionless coefficient (recall that we are working in units where $z_\star \equiv 1$).
\R\\
\emph{Bulk $\phi$ Solutions}\\

The action for a real massless scalar field in $D$ space-time dimensions is
\begin{equation}
S_{\phi} =\f{1}{2} \int d^{D} x \sqrt{-g}\del_\m\phi\del^\m\phi 
\end{equation}
whose variation with respect to $\phi$ yields the Klein-Gordon equation on this space-time,
\begin{equation}
 \Box\phi = \(-\d_t^2  + \del^2 \)  \phi = 0
\end{equation}
where
\begin{equation}
\del^2=\del_{\bf x}^2 + e^{2z} \f{\d^2}{\d \th^2} +\f{\d^2}{\d z^2}-\f{\d}{\d z}
\end{equation}
and $\del_{\bf x}^2$ is the Minkowski-space Laplacian.  The positive frequency modes, $u_{\bf i}$, are

\begin{equation}\label{u_i, 3}
u_{\bf i}=A_{\bf i} e^{- i (\o t - {\bf p}\cdot {\bf x} - n\th )} Z_{n,k}(z)\,,
\end{equation}
where $n \in \mathbb{Z}$ and $A_{\bf i}$ is a normalization constant. $Z_{n,k}$ satisfies
\begin{equation}\label{horn_zeqn}
Z_{n,k}''(z) - Z_{n,k}'(z) + \(\o^2 - p^2 - n^2 e^{2z}\)Z_{n,k}(z)=0 \,, 
\end{equation}
where $p=\abs{{\bf p}}$ is the momentum in the Minkowski directions. We find that the boundary conditions and normalizability imply a real $k>0$, defined by the dispersion relation
\begin{equation}\label{dispersion_relation}
\o =   \sqrt{p^2 +k^2 +\f{1}{4}}
\end{equation}
%


To make the problem more tractable, we regulate the infinite spatial volume of the horn by truncating the space at $z=z_L$, where $z_L < z_b$, and impose there a Dirichlet boundary condition on the field\textsuperscript{\footnotemark[4]}\footnotetext[4]{This makes the modes discrete and normalizable. The choice of Dirchlet boundary condition is arbitrary and should have no bearing on the end result.}. At the appropriate place in the calculation we take the limit $z_L\to -\infty$ to recover the full horn spacetime. For the $n\neq0$ modes we find (see Appendix \ref{solutions_horn})
\begin{equation}
\label{horn_soln_nneq0_leftright} 
Z_{n\neq0,k}=
\begin{cases}
e^{z/2}\[I_{-i k}\(\abs{n} e^{z_L}\) I_{ik}\(\abs{n} e^{z}\) -I_{ik}\(\abs{n} e^{z_L}\) I_{-ik}\(\abs{n} e^{z}\)\]~&\mbox{when}\quad z_L \leq z \leq z_b\,,\\
e^{z/2} K_{ik}\(\abs{n} e^{z}\) &\mbox{when}\quad  z_b \leq z\,.\end{cases}
\end{equation}

The $n=0$ modes do not contribute to the Casimir force, as we show  explicitly in Appendix \ref{Horn_gen_neq0};
however this can be understood from the following argument. From \eqref{horn_zeqn} we see that these solutions behave as $e^{z/2}\sin{k(z-z_{L,R})}$, and the energy density \emph{per unit $z$}  in these modes is proportional to $\sqrt{\abs{g}}Z_{0,k}^2 = \sin^2{k(z-z_{L,R})}$, where  $z_{L}$ and $z_R$ are simply there to temporarily regulate the size of the space in the $z$-direction.  This is the same form as the energy density of scalar modes in a 1-dimensional box; as the box size goes to infinity we know there is no Casimir force due to translation invariance. Thus the $n=0$ modes on the horn do not contribute to the Casimir force, essentially, because they don't ``feel" that translation invariance of the manifold is broken.

For all $n$, the spectrum of $k$ are determined by the boundary conditions at $z_b$:
\begin{equation}
0 = \begin{cases} Z_{n,k}(z_b) &\mbox{(Dirichlet)}\,,  \\ \mbox{or}&  \\ Z_{n,k}'(z_b) &\mbox{(Neumann)}\,.
\end{cases}\\
\end{equation}

Finally, the field $\phi$ may be expanded in terms of the solutions as 
\begin{equation}
\phi=\sum_{\bf i} a_{\bf i} u_{\bf i} + a^\dagger_{\bf i}u^*_{\bf i}\,,
\end{equation}
where $a^\dagger_{\bf i}$ and $a_{\bf i}$ are the creation and annihilation operators of modes labeled by the set of quantum numbers, $\bf i$. As usual, the vacuum state is defined as $a_{\bf i} \ket{0}=0$ and the operators satisfy the  commutation relations, $[a_{\bf i},a_{\bf j}]=0=[a^\dagger_{\bf i},a^\dagger_{\bf j}]$, and $[a_{\bf i}, a^\dagger_{\bf j}]=\de_{\bf ij}$. The positive-energy eigenfunctions are normalized using the Klein-Gordon norm (assuming a discrete spectrum):
\begin{align}
(u_{{\bf i}}, u_{{\bf j}}) &\equiv i \int \sqrt{-g}~ d \Sigma n^\m ( u_{\bf i}^* \del_\m  u_{\bf j} -  u_{\bf j} \del_\m  u_{\bf i}^*    )\equiv\de_{\bf ij}
\end{align}
where $\Sigma$ is a spacelike hypersurface, and $n^\m$ is a unit timelike vector normal to it.

\sssec{Procedure for Calculation of Casimir Energy on the Horn}\label{horn_technical}
~\\
\emph{Contour Integral Representation of Sum}\label{gen_contour_rep}\\

As we generally have no explicit expression for the $\{k\}$, we will calculate \eqref{zeta_fn_reg_E_0} in part using a contour integral representation, as elucidated in \cite{Bordag:2009zz}.  To this end, the following will be of use for both the $D=$ 2+1 and 5+1 models (i.e. $m=0$ and $m=3$):
\begin{align}\label{contour_rep}
\sum_{\{k\}}\(k^2 + \f{1}{4}\)^{\f{\(m+1\)}{2}-s} =\f{1}{2\p i}\oint_\g dk \(k^2 + \f{1}{4}\)^{\f{\(m+1\)}{2}-s}\f{\d}{\d k} \ln \Delta_n(k)
\end{align}
where $\g$ is a counter-clockwise contour that encloses the entire spectrum (the positive real axis) and $\Delta_n(k)$ are appropriate mode-generating functions whose roots correspond to our spectrum of $k$ (for details see Appendix \ref{contour_details}, or e.g. references  \cite{Bordag:2009zz} or \cite{Fursaev:2011zz}).
With properly chosen generating functions, one can show that the contour may be deformed to give
\begin{equation}\label{rotated}
\sum_{\{k\}}\(k^2 + \f{1}{4}\)^{\f{\(m+1\)}{2}-s}=-\f{\cos{\p \(\f{m}{2}-s\)}}{\p}\int_{1/2}^{\infty} dk \(k^2 - \f{1}{4}\)^{\f{\(m+1\)}{2}-s}\f{\d}{\d k} \ln \Delta_n(i k)
\end{equation}

We have derived a set of mode generating functions for $n\neq0$ and $n=0$ separately (Appendix \ref{contour_details}), and have shown that the $n=0$ modes are independent of $z_b$, and so \emph{we omit $n=0$ from any sums we encounter}.  We find the mode generating functions for both sides of $z_b$ to effectively combine to give the log of the total generating functions
\begin{equation}
\label{logDeltaBoth}
\ln{\Delta_n(i k)}=
\begin{cases}
\ln{\[I K\]} &\mbox{(Dirichlet)},\\
\ln{\[\f{1}{4}I K + \f{1}{2}x_b\(I' K + I K'\) + x_b^2 I'K'\]}&\mbox{(Neumann),}
\end{cases}
\end{equation}
where, for brevity, we denote $I_k\(\abs{n} e^{z_b}\)$ by $I$ and $K_k\(\abs{n} e^{z_b}\)$ by $K$,  and $'$ denotes a derivative with respect to the argument of the Bessel function. Any $z_b$-independent terms have been omitted as they do not contribute to the Casimir force.
\R
\emph{General Analysis of $E_0(s)$}\\

The divergent piece from $E_0(s)$ needs to be separated out so that its physically relevant  part may be revealed through an analytic continuation in $s$, else the divergent quantities need to be explicitly absorbed through some renormalization of parameters in the full theory.  Because the summand is even in $n$ and $n=0$ isn't counted, we may take
\begin{equation}
\sum_n \to 2 \sum_{n=1}^\infty
\end{equation}
so that in general we have
\begin{equation}\label{E_0_gen}
E_0(s) \propto \sum_{n=1}^\infty \int_{1/2}^{\infty} dk \(k^2 - \f{1}{4}\)^{\f{\(m+1\)}{2}-s}\f{\d}{\d k} \ln{\Delta_n(i k)}
\end{equation}

Given \eqref{logDeltaBoth}, an analytic continuation of \eqref{E_0_gen} in $s$ is impossible in its current form, so we approach it using a uniform asymptotic expansion, closely following the procedure in \cite{Bordag:2009zz}.  We know the divergences occur at large $k$ and $n$, but the idea is to isolate their asymptotic behavior, taking them to infinity simultaneously while keeping their ratio fixed. If this asymptotic behavior can be understood analytically, it is then straightforward to handle the divergent contributions so that the remaining finite part may be calculated numerically. To this end, we decompose $E_0(s)$ into a sum of its asymptotic and finite part:
\begin{equation}\label{sigma}
E_0(s) \equiv E_0^{\text{as}}(s)+E_0^{\text{\text{fin}}}
\end{equation}
We change variables, defining
\begin{equation}
 x_b \equiv \abs{n}e^{z_b}
\end{equation}
and
\begin{equation}
y \equiv \f{k}{x_b}.
\end{equation}
%
Our sum/integral therefore becomes
 \begin{equation}\label{E_0_gen_2}
E_0(s)\propto\sum_{n=1}^\infty x_b^{m+1-2s}\int_{(2 x_b)^{-1}}^{\infty} dy \(y^2 - \f{1}{4 x_b^{2}}\)^{\f{\(m+1\)}{2}-s}\f{\d}{\d y} \ln{\Delta_n(i y x_b)}\,.
\end{equation}
From the uniform asymptotic behavior of the two modified Bessel functions (see e.g. \cite{abramowitz+stegun}), we define the asymptotic part of the log of the generating function as
\begin{align}
\widetilde{\ln}{\[\Delta_n(i y x_b)\]} &\equiv \ln{\[\f{1}{2x_b \sqrt{y^2+1}}\]} +\sum_{j=1}^{m+2} \f{f_{j}(y^{-1})}{(y x_b)^{j}}~~~~~\text{(Dirichlet)}\label{log_KI_expansion}\\
&\equiv
\ln{\[-\f{x_b}{2}\sqrt{y^2+1}\]} +\sum_{j=1}^{m+2} \f{g_{j}(y^{-1})}{\(y x_b\)^{j}}~~~~~\text{(Neumann),}
\label{log_KpIp_expansion}
\end{align}
where the $f_{j}(z)$ and $g_{j}(z)$ are rational functions of $z$, given in Appendix \ref{Bessel_asymptotics}. 
%
We find it most convenient to capture the divergences of \eqref{E_0_gen_2} by defining\textsuperscript{\footnotemark[5]}\footnotetext[5]{This turns out to be easier than doing an expansion in inverse powers of $x_b$ of the entire integrand.}
\begin{align}\label{E_0_as}
E_0^{\text{as}}\(s\) \propto  \sum_{n=1}^\infty & x_b^{m+1-2s}\int_{(2 x_b)^{-1}}^{\infty} dy \(y^2 - \f{1}{4 x_b^{2}}\)^{\f{\(m+1\)}{2}-s} \f{\d}{\d y} \widetilde{\ln}{\[I_{yx_b}\(x_b\)K_{yx_b}\(x_b\)\]}.
\end{align}
for the Dirichlet case, and similarly for Neumann. We will see that the integral may be performed analytically, however the sum must be done numerically and so an analytic continuation again becomes challenging. We will then find it convenient to further decompose $E_0^{\text{as}}(s)$ into divergent and ``remainder" parts, $E_0^{\text{div}}(s)$ and $E_0^{\text{rem}}$. In summary, the total energy (or energy density) of the system will be given by

\begin{align}\label{general_E0}
E &= E^{~}_{\text{ten}} + E_0^{\text{div}}(s)+E_0^{\text{rem}}+E_0^{\text{fin}}~~~~~~&(D=2+1)\\
\r&=\r^{~}_{\text{ten}} + \r_0^{\text{div}}(s)+\r_0^{\text{rem}}+\r_0^{\text{fin}} + \r_\text{0, SM}&(D=5+1)
\end{align}

\ssec{$D=2+1$}\label{section:2+1D}


Here there are no Minkowski dimensions, so ${\bf p}=0$. From \eqref{zeta_fn_reg_E_0} and \eqref{rotated} we have 

\begin{align}
E_0(s)
=-\f{\m^{2s}}{\p}\cos{\p s} \sum_{n=1}^\infty \int_{1/2}^{\infty} dk \(k^2 - \f{1}{4}\)^{1/2-s}\f{\d}{\d k} \ln{\Delta_k(ik)}\label{E0_3D_horn}
\end{align}
where $\ln \Delta_n(ik)$ is defined by \eqref{logDeltaBoth} in both the Dirichlet and Neumann cases.


\sssec{Dirichlet Condition}\label{2+1_Horn_D}
Many of the details of this section are included in Appendix \ref{2+1_Horn_details} and maybe be easily generalized to the other cases. Considering \eqref{log_KI_expansion} and \eqref{E_0_as}, we define the asymptotic part of the energy as
\begin{equation}\label{E0as_3D}
E_0^{\text{as}}(s)= -\f{\m^{2s}}{\p}\cos{\p s} \sum_{n=1}^\infty x_b^{1-2s}\int_{(2 x_b)^{-1}}^{\infty} dy \(y^2 - \f{1}{4 x_b^{2}}\)^{1/2-s}\f{\d}{\d y}\widetilde{\ln}{\[I_{yx_b}\(x_b\)K_{yx_b}\(x_b\)\]} 
\end{equation}

After analytically continuing towards $s=0$ we find 
\begin{equation}
E_0^{\text{as}}(s)=-\f{(2\m)^{2s}}{4}\cos{\p s} \sum_{n=1}^\infty\f{1+16x_b^2 + (49-10s) x_b^4 + 64 x_b^6}{\(1+4 x_b^2\)^{5/2+s}}
\end{equation}
As $x_b \propto n$, the divergences here will come from the terms in the summand that go asymptotically as $x_b$ or $x_b^{-1}$; we extract these by defining
\begin{equation}
E_0^{\text{as}}(s)\equiv E_0^{\text{\text{div}}}(s)+E_0^{\text{\text{rem}}}
\end{equation}
where we have expanded in inverse powers of $x_b$ to obtain
\begin{align}\label{E0div_3D}
E_0^{\text{div}}(s) &= -\m^{2s}\cos{\p s} \sum_{n=1}^\infty \( \f{1}{2} x_b^{1-2s}+ \f{9-26s}{128}x_b^{-1-2s}\)\notag\\
&=  \f{e^{z_b}}{24} +\f{9}{128}e^{-z_b}\(\f{13}{9} - \g - \f{1}{2s}  - \ln{\[\m e^{-z_b}\]} \)
\end{align}
and (evaluated with $s=0$)
\begin{align}\label{E0rem_3D}
E_0^{\text{\text{rem}}}= \sum_{n=1}^\infty\[-\f{1+16x_b^2 + 49 x_b^4 + 64 x_b^6}{4\(1+4 x_b^2\)^{5/2}} + \( \f{1}{2} x_b+ \f{9}{128}x_b^{-1}\)\]
\end{align}
Asymptotically, we find 
\begin{equation}
\label{E0_rem_horn}
E_0^{\text{\text{rem}}} \sim
\begin{cases}
 -\f{7+36 \ln{2}}{256}\times e^{-z_b}  & \mbox{as}\quad  z_b\to -\infty \,,\\
 -\f{23}{1024}\zeta(3)\times e^{-3z_b} & \mbox{as}\quad  z_b \to +\infty \,.
\end{cases}
\end{equation}
where we have used a lowest order Euler-Maclauren expansion in the first line, and a Taylor expansion in $x_b^{-1}$ in the second. After setting $s=0$, integrating by parts\textsuperscript{\footnotemark[6]}\footnotetext[6]{This is advisable for numerical purposes. 
}, and returning to the variables $k$ and $x_b$ one arrives at
 \begin{equation}\label{E0fin_3D}
E_0^{\text{fin}}=\f{1}{\p} \sum_{n=1}^\infty \int_{1/2}^{\infty} dk ~\f{k}{\(k^2 - \f{1}{4}\)^{1/2}} \(\ln \[I_{k}\(x_b\)K_{k}\(x_b\)\] - \widetilde{\ln}{\[I_{k}\(x_b\)K_{k}\(x_b\)\]}\)
\end{equation}
Asymptotically, we find
\begin{equation}
\label{E0_fin_horn}
E_0^{\text{\text{fin}}} \sim
\begin{cases}
2.4\times 10^{-3}\times e^{-z_b}  & \mbox{as}\quad  z_b\to -\infty \,,\\
-1.4\times 10^{-3}\times e^{-3z_b} & \mbox{as}\quad  z_b \to +\infty \,.
\end{cases}
\end{equation}

\begin{figure}
  \begin{center}
    \includegraphics[scale=.5]{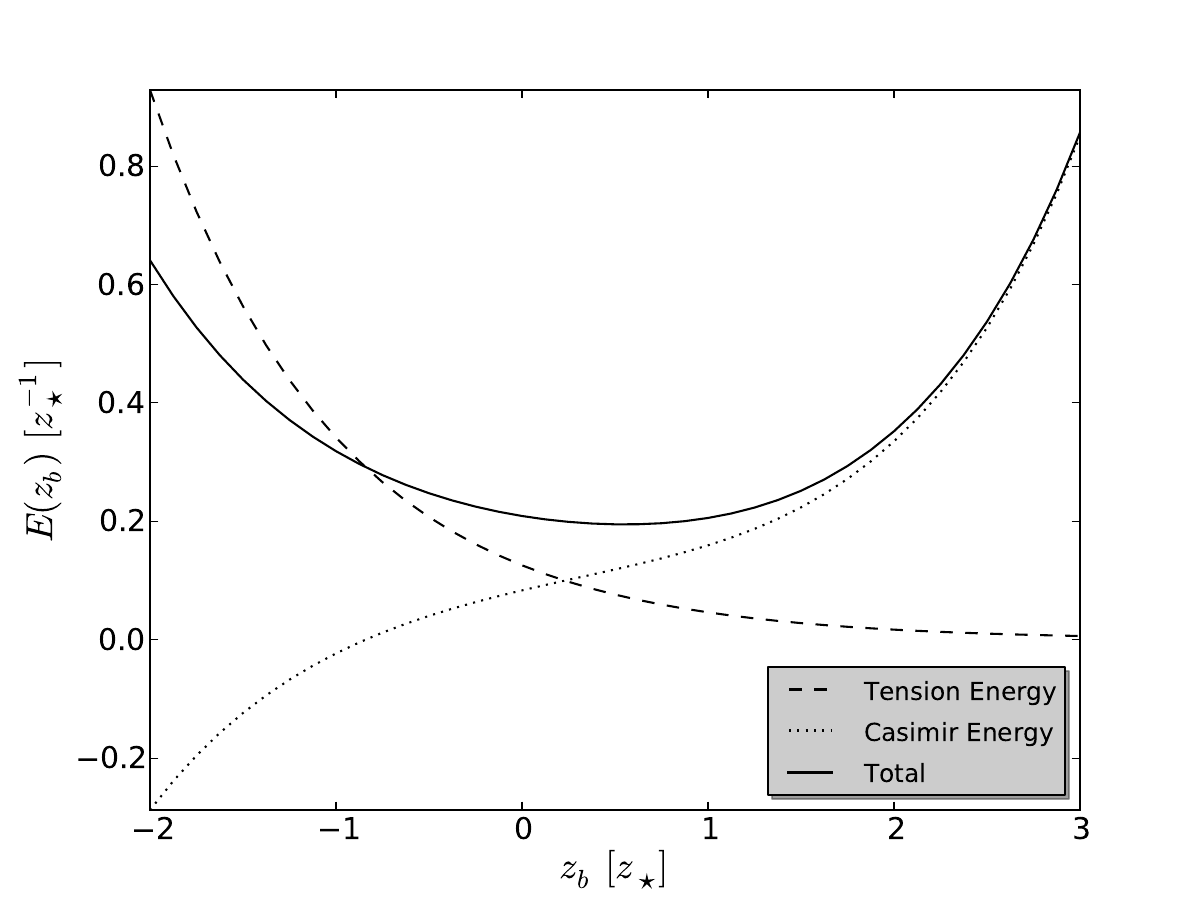}
  \end{center}
\caption{In $D=2+1$ with a Dirichlet boundary condition on $\phi$, various potential contributions as a function of $z_b$. For illustration we have chosen $\s_\text{ren}=0.02~[z_\star^{-2}]$.}
\label{3DHornDirichlet}
\end{figure}

We have learned that $E_0^\text{div}$ contains an explicit $1/s$ pole which is proportional to $e^{-z_b}$; we have also verified this using a complementary heat kernel analysis (see Appendix \ref{heat_appendix}). The presence of the pole and the arbitrary scale $\m$ is indicative of the fact that the Casimir energy cannot be measured in isolation; it is only a contribution to a total energy that is itself finite \cite{Blau:1988kv}. One must identify a parameter that is to be renormalized --  since this offending term is proportional to the brane length, a renormalization of the brane tension will suffice, with $\mu$ setting the renormalization scale. In so doing, we choose to go beyond minimal subtraction and absorb finite corrections into the tension as well, hence we find the asymptotic behaviors  (see Appendix \ref{2+1_Horn_details})
\begin{equation}
E(z_b) \sim
\begin{cases}
2\p z_\star  e^{-z_b/z_\star}\times \s_{\text{ren}} & \mbox{as}\quad  z_b\to -\infty \,,\\
 \f{1}{24}  \f{e^{z_b/z_\star}}{z_\star} & \mbox{as}\quad  z_b \to +\infty \,.
\end{cases}
\end{equation}
where we have restored the horn length scale, $z_\star$. We find that the total energy is unbounded from above as $z_b \to +\infty$ and likewise for $z_b \to -\infty$ as long as $\s_{\text{ren}} >0$, and therefore a global minimum exists. $E(z_b)$ is plotted in Figure \ref{3DHornDirichlet}.

\sssec{Neumann Condition}
Here we repeat the analysis of the previous section, only now using the Neumann specification in \eqref{logDeltaBoth}. Altogether, we find the asymptotic behaviors
\begin{equation}
E(z_b) \sim
\begin{cases}
2\p z_\star  e^{-z_b/z_\star}\times \s_{\text{ren}} & \mbox{as}\quad  z_b\to -\infty \,,\\
 -\f{1}{24}  \f{e^{z_b/z_\star}}{z_\star} & \mbox{as}\quad  z_b \to +\infty \,.
\end{cases}
\end{equation}

\begin{figure}
  \begin{center}
    \includegraphics[scale=.5]{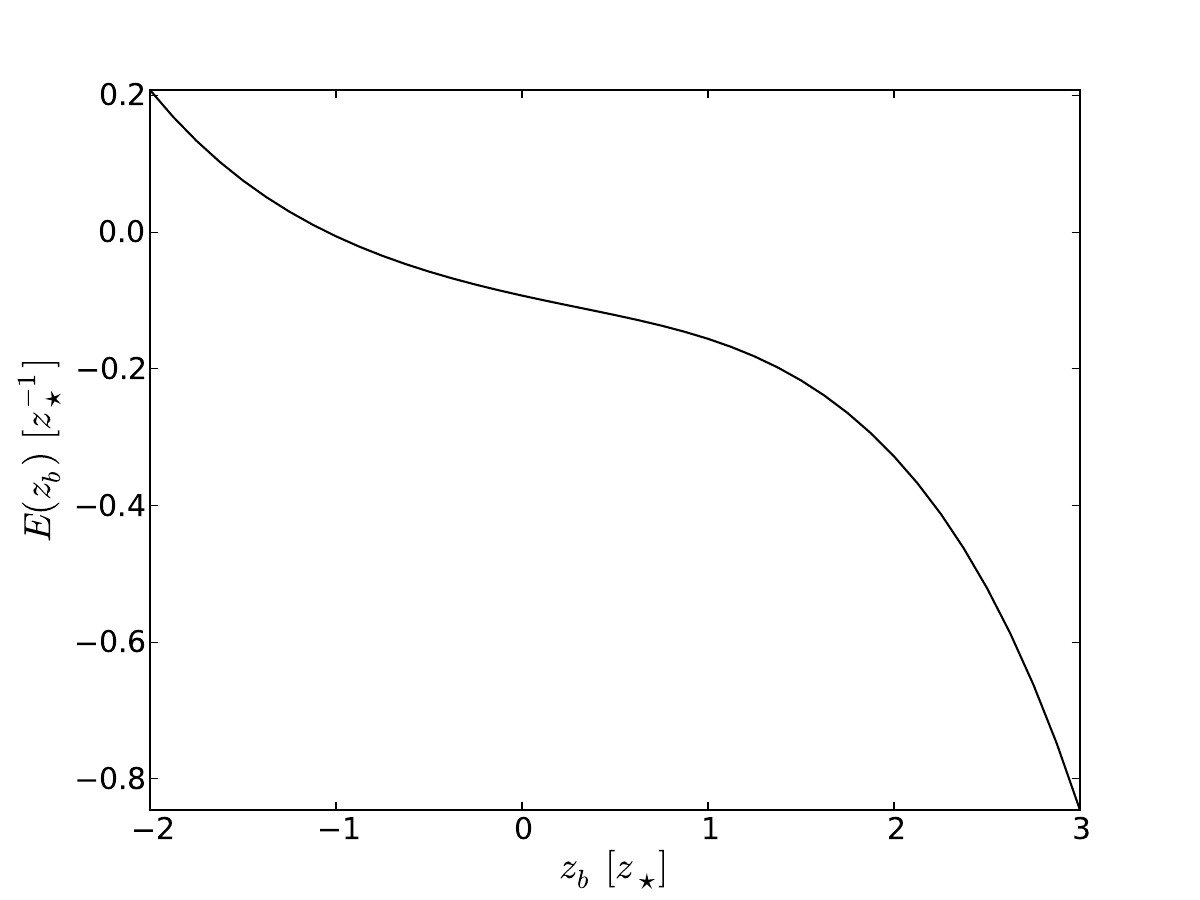}
  \end{center}
\caption{In $D=2+1$ with a Neumann boundary condition, the total energy as function of $z_b$. Here we have chosen $\s_\text{ren}=0$.}
\label{3D_Horn_Neumann_Plot}
\end{figure}

We find that the Casimir energy is unbounded from below for $z_b \to +\infty$ (the sign of the energy is flipped with respect to the Dirichlet case, in this limit), and brane tension cannot compensate for this,  therefore no stabilizing mechanism exists in this case.   $E(z_b)$ is plotted in Figure \ref{3D_Horn_Neumann_Plot} without brane tension.

\ssec{$D=5+1$}\label{section:5+1D}
Here the previous analysis is repeated, adding three Minkowksi spatial dimensions, i.e. $m=3$. For ease of normalization we compactify these dimensions using a torus of fundamental side length, $V_\text{M}^{1/3}$, which is taken to infinity to recover a true global Minkowski space. In this limit, the Minkowski momentum, ${\bf p}$, becomes continuous.  The sum over the components of $\bf{p}$ turns into an integral as 

\begin{equation}
\sum_{\bf{p}} = \sum_{\bf{p}}  \f{\De p_x}{2\p/V_\text{M}^{1/3}}\f{\De p_y}{2\p/V_\text{M}^{1/3}}\f{\De p_z}{2\p/V_\text{M}^{1/3}} \to \f{V_\text{M}}{(2\p)^3}\int dp_x dp_y dp_z = \f{V_\text{M}}{2\p^2}\int dp p^2
\end{equation}

From \eqref{zeta_fn_reg_E_0} we now have the effective 4D vacuum energy

%
\begin{equation}
\r_0\(s\)= \f{\m^{2s} }{2\p^2}\sum_{n=1}^{\infty}\sum_{\{k\}}\int dp p^2 \(p^2 + k^2 +\f{1}{4}\)^{1/2-s}
\end{equation}

We compute the $p$-integral by first performing a Mellin transform,
\begin{equation}
\(p^2 + k^2 +\f{1}{4}\)^{1/2-s} = \int_0^\infty dt \f{t^{s-3/2}}{\Gamma(s-1/2)}e^{-t \(p^2 + k^2 +\f{1}{4}\)}
\end{equation}
which allows us to directly integrate over $p$, yielding
\begin{align}
\r_0\(s\)
&= \f{\m^{2s} }{8\p^{3/2}}\f{\Gamma(s-2)}{\Gamma(s-1/2)}\sum_{n=1}^{\infty}\sum_{\{k\}}   \( k^2 +\f{1}{4}\)^{2-s}
\end{align}
We now represent the sum over $\{k\}$ as a contour integral using \eqref{rotated}, giving
\begin{align}
\r_0\(s\)
&=\f{\m^{2s} }{8\p^{5/2}}\f{\Gamma(s-2)\sin{\p s}}{\Gamma(s-1/2)} \sum_{n=1}^\infty \int_{1/2}^\infty dk   \( k^2 -\f{1}{4}\)^{2-s} \f{\d}{\d k} \ln \De_n(ik)\notag\\
&= -\f{\m^{2s} }{32\p^{2}}\(1+s\(\f{3}{2}-\g -\psi(\sfrac{-1}{2})\)\)  \sum_{n=1}^\infty \int_{1/2}^\infty dk   \( k^2 -\f{1}{4}\)^{2-s} \f{\d}{\d k} \ln \De_n(ik)
\end{align}
where $\psi$ is the Digamma function, and we have omitted terms higher order in $s$ as they vanish in the $s\to 0$ limit. 

\sssec{Dirichlet Condition}
Considering \eqref{log_KI_expansion} and \eqref{E_0_as} we define
\begin{align}
\r_0^{\text{as}}(s)=-\f{\m^{2s} }{32\p^{2}}&\(1+s\(\f{3}{2}-\g -\psi(\sfrac{-1}{2})\)\)  \sum_{n=1}^\infty  x_b^{4-2s}\int_{(2 x_b)^{-1}}^{\infty} dy \(y^2 - \f{1}{4 x_b^{2}}\)^{2-s}\notag\\
&~~~~~~~~~~~~~~~\times\f{\d}{\d y} \widetilde{\ln}{\[I_{yx_b}\(x_b\)K_{yx_b}\(x_b\)\]} 
\end{align}
The integral may be performed analytically, yielding a cumbersome expression which we omit for brevity.
Here the divergences come from terms in the summand that behave asymptotically as $x_b^4, x_b^2$ or $x_b^0$.  We isolate those terms in (having already performed the sum on $n$)
\begin{align}\label{E_0_div_6D_Dirichlet}
\r_0^{\text{div}}(s)=\f{1}{2048\p^2}\(-\f{47}{12}+\g-\f{1}{s}- 2\ln{\p\m e^{-z_b}} + \psi(\sfrac{-1}{2})\) -\f{3}{64\p^2}\zeta'(-2)e^{2z_b}+\f{1}{32\p^2}\zeta'(-4)e^{4z_b}
\end{align}
From this we numerically obtain
\begin{equation}
\r_0^{\text{rem}}= \lim_{s\to0}\(\r_0^{\text{as}}(s)-\r_0^{\text{div}}(s)\)
\end{equation}
which asymptotically behaves as
\begin{equation}
\r_0^{\text{rem}}\sim
\begin{cases}
-1.5 \times 10^{-4} \times e^{-z_b} & \mbox{as}\quad  z_b\to -\infty \,,\\
6.8\times 10^{-5}\times e^{-2z_b}& \mbox{as}\quad  z_b \to +\infty \,.
\end{cases}
\end{equation}
Lastly, we return to the variables $k, x_b$ and integrate by parts to give
 \begin{align}\label{E_0_fin_6D_Dirichlet}
\r_0^{\text{fin}}=\f{1}{8\p^{2}}\sum_{n=1}^\infty& \int_{1/2}^{\infty} dk~k \(k^2 - \f{1}{4}\) \(\ln \[I_{k}\(x_b\)K_{k}\(x_b\)\] - \widetilde{\ln}{\[I_{k}\(x_b\)K_{k}\(x_b\)\]}\),
\end{align}
Asymptotically, one may show
\begin{equation}
\r_0^{\text{fin}}\sim
\begin{cases}
-2.0 \times 10^{-5} \times e^{-z_b} & \mbox{as}\quad  z_b\to -\infty \,,\\
4.4\times 10^{-5}\times e^{-2z_b}& \mbox{as}\quad  z_b \to +\infty \,.
\end{cases}
\end{equation}

\begin{figure}
  \begin{center}
    \includegraphics[scale=.5]{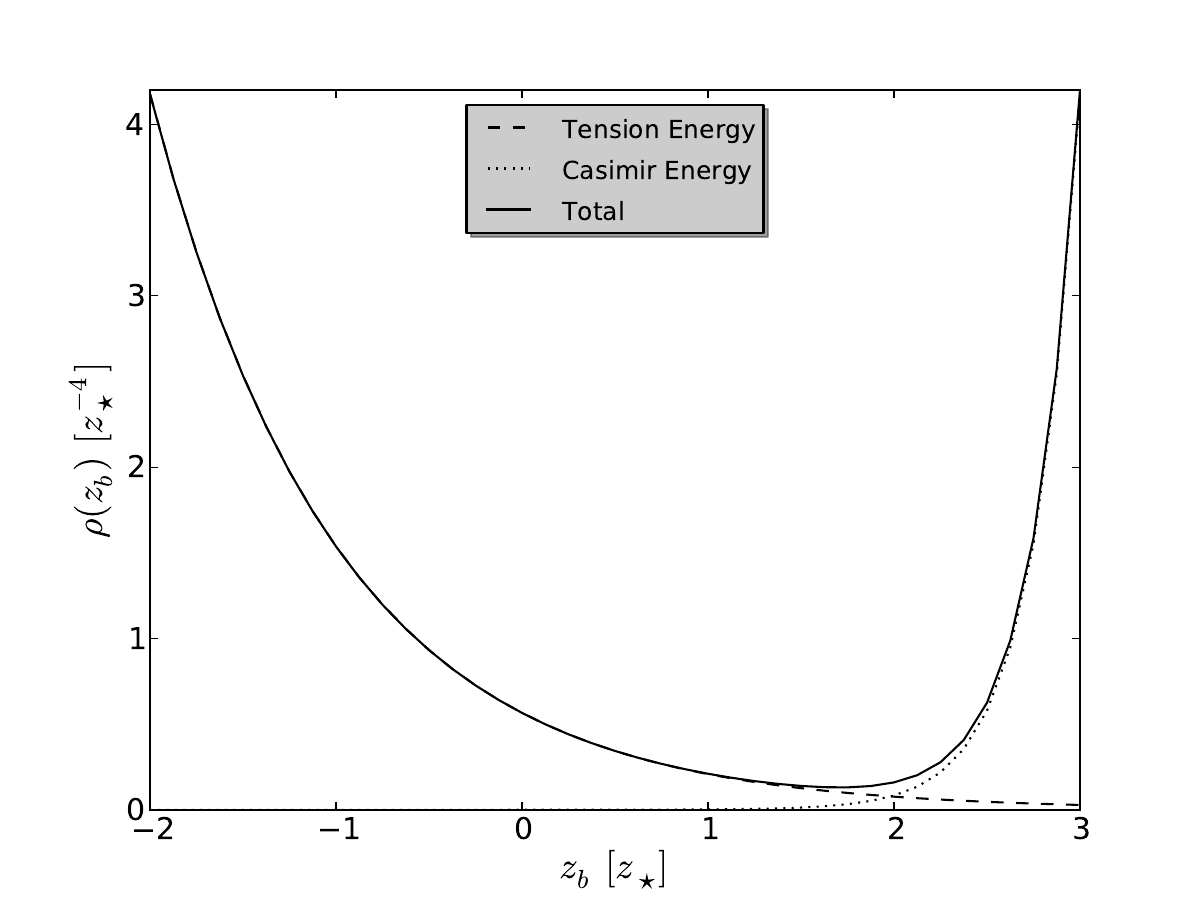}
  \end{center}
\caption{For $D=5+1$ with the Dirichlet condition on $\phi$, the total 4D energy density in units of the horn curvature scale ($z_\star$), as a function of the brane position, $z_b$. For illustration, we have taken the brane tension, $\s_\text{ren} =0.09$ [$z_\star^{-5}$], and $\k_{SM}=0$.}
\label{5plus1_Dirichlet}
\end{figure}

In contrast to the 2+1 case, the pole in $E_0^{\text{div}}(s)$ is a constant and is simply discarded. Altogether we have the asymptotic behaviors
\begin{equation}
\r(z_b)\sim
\begin{cases}
2\p z_\star e^{-z_b/z_\star}\times \s_\text{ren} & \mbox{as}\quad  z_b\to -\infty \,,\\
\f{1}{16\p^4 z_\star^4}
\(\f{1}{2}\p^2 \zeta'(-4) + \k_{SM}\)\times{e^{4z_b/z_\star}}& \mbox{as}\quad  z_b \to +\infty \,.
\end{cases}
\end{equation}
where finite quantum corrections have been absorbed by $\s_\text{ren}$, and  $1/2\p^2 \zeta'(-4) \approx 0.04$. As  in the $D=2+1$ case, here there is a global minimum, assuming $\s_\text{ren} >0$ and assuming the value of $\k_{SM}$ doesn't alter these results (see Figure \ref{5plus1_Dirichlet}).  For $\s ={\cal O}(1)~[z_\star^{-5}]$, the effective mass for the brane's position modulus is ${\cal O}(z_\star^{-1})$.

\sssec{Neumann Condition}
Here we proceed as in the previous section, only using \eqref{log_KpIp_expansion}. Altogether we find the asymptotic behaviors
\begin{equation}
\r(z_b) 
\sim
\begin{cases}
 2\p z_\star e^{-z_b/z_\star}\times \s_\text{ren}
 & \mbox{as}\quad  z_b\to -\infty \,,\\
 \f{1}{16\p^4 z_\star^4}
\(-\f{1}{2}\p^2 \zeta'(-4) + \k_{SM}\) \times{e^{4z_b/z_\star}}
& \mbox{as}\quad  z_b \to +\infty \,.
\end{cases}
\end{equation}

\begin{figure}
  \begin{center}
    \includegraphics[scale=.5]{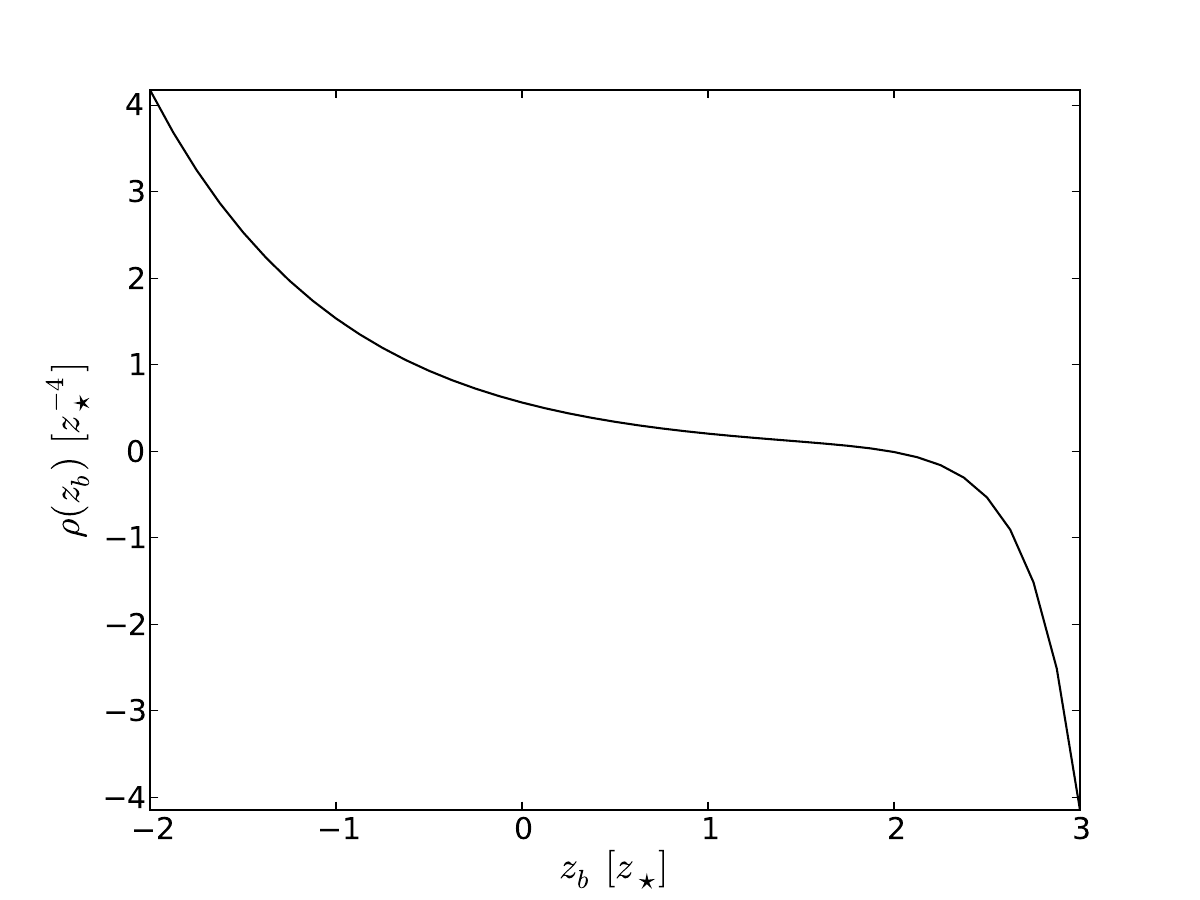}
  \end{center}
\caption{For $D=5+1$ with the Neumann condition on $\phi$, the total 4D energy density in units of the horn curvature scale ($z_\star$), as a function of the brane position, $z_b$. Here we have chosen $\s_\text{ren} =0$, and $\k_{SM}=0$.}
\label{6DHornNeumann}
\end{figure}

\begin{figure}
 \begin{center}
    \includegraphics[scale=1]{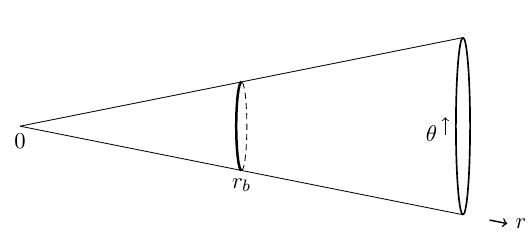}
\end{center}
\caption{A partial embedding diagram of the cone. The brane is pictured at coordinate $ r_b$. As opposed to the horn, this manifold is geometrically flat and is infinite in extent only in one direction.} 
\label{fig_cone}
\end{figure}

Once again, the sign of the vacuum energy in $\phi$ is flipped with respect to the Dirichlet case in the $z_b \to \infty$ limit. It is thus unbounded from below, and no value of brane tension can create a stable minimum, however a sufficiently large and positive $\k_{SM}$ would accomplish this (see Figure \ref{6DHornNeumann}). 

\sec{Euclidean Cone}\label{cone}

\ssec{Preliminaries}\label{section:prelims_cone}
\sssec{The Model, Energy Contributions, and Bulk $\phi$ Solutions}
Once more we work in general spacetime dimension $D=m+2+1$, where $m$ indicates the number of Minkowski spatial dimensions. That is, the topology of the full spacetime manifold is ${\cal M}^{m+1}\times \(\mathbb{R}^2-\{0\}\)$. The line element can be written in coordinates such that
\begin{equation*}
ds^2 = \e_{\m\n}^{(m+1)}dx^\m dx^\n  + dr^2 + r^2 d\th^2 
\end{equation*}
where we identify $\th \leftrightarrow \th +2\p\(1-\de\)$,  i.e. the conical space is defined by the deficit angle $2\p \de$. Again, a 1-brane (4-brane) is wrapped around this space in $D=2+1$ ($D=5+1$) and resides at coordinate distance $r_b$ away form the cone tip, as illustrated in Figure \ref{fig_cone}. As opposed to the horn, the cone is flat, having no intrinsic length scale associated with it.
\\
\\
\emph{Energy Contributions}\\

Here, a non-zero brane tension ($\s$) contributes an energy given by
\begin{equation}
E^{~}_{\text{ten}}=\int d^{1+m}x \sqrt{\abs{\g}}  \s = 2\p \(1-\de\) r_b V_{M}  \s
\end{equation}
where $\g_{\m\n}$ is the induced spatial metric on the brane, while the elastic energy characterized by the extrinsic curvature of the brane, $K_{ab}$ (see Appendix \ref{heat_cone}) is
\begin{equation}
E^{~}_{\text{curv}}=\int d^{1+m}x \sqrt{\abs{\g}} \(h_1K^{2}+h_2K_{ab}K_{ab}+\dots\) = \f{2\p \(1-\de\)}{ r_b} V_{M} \(h_1+h_2+\dots\)
\end{equation}
where the $h_i$ are parameters describing the energy cost for deforming the brane within the manifold, and $\dots$ represents other possible scalars\textsuperscript{\footnotemark[7]}\footnotetext[7]{Coupling to the Riemann curvature tensor is trivial as it is zero for this geometry.} that are built from terms higher order in $K_{ab}$.   Note that the effect of each of these terms is completely indistinguishable from one another, at least as far as the total energy of the system is concerned. Without loss of generality, then, we shall define $h_1+h_2+\dots\equiv h$.  As opposed to the horn, the tension and elastic energies appear with markedly different dependences on the brane location. Interestingly, it is already apparent that these two energies are sufficient to stabilize the brane under the physically reasonable assumption that $\s, h > 0$.

The zero-point energy in the bulk $\phi$ is again zeta-function regularized and is given by \eqref{zeta_fn_reg_E_0}. As we will see below, a check on our results is available in $D=2+1$ when $\de=0$ as this is simply the Casimir effect of the 1-sphere.
In $D=5+1$ we parametrize the vacuum energy of brane-dwelling Standard Model fields, from dimensional considerations, by
\begin{equation}
\r_\text{0, SM} \simeq   \f{\k_{SM}}{\big(2\p\(1-\de\) r_b\big)^{4}}
\end{equation}
where $\k_{SM}$ is a dimensionless coefficient. 
\R\\
\emph{Bulk $\phi$ Solutions}\\

Again, the massless bulk scalar field obeys the Klein-Gordon equation,
\begin{equation}
 \Box\phi = \(-\d_t^2  + \del^2 \)  \phi = 0
\end{equation}
where
\begin{equation}
\del^2=\del_{\bf x}^2 + \f{\d^2}{\d r^2} + \f{1}{r}\f{\d}{\d r}+\f{1}{r^2}\f{\d^2}{\d \th^2}
\end{equation}
and $\del_{\bf x}^2$ is the Minkowski-space Laplacian.  The positive frequency modes, $u_{\bf i}$, are found to be

\begin{equation}
u_{\bf i}=A_{\bf i} e^{- i (\o t - {\bf p}\cdot {\bf x} - \n\th )} R_{n,k}(r)
\end{equation}
where $\n\equiv n/(1-\de)$, $n \in \mathbb{Z}$, $A_{\bf i}$ is a normalization constant, and $R_{n,k}$ satisfies Bessel's equation
\begin{equation}
R_{n,k}''(r) +\f{1}{r}R_{n,k}'(r)+\(k^2-\f{\n^2}{r^2}\)R_{n,k}(r)=0
\end{equation}
whose solutions may be written as a linear combination of the Bessel functions of the first and second kind, $J_\n(k r)$ and $N_\n(k r)$, or the Hankel functions $H^{(1)}_\n(k r)$ and $H^{(2)}_\n(k r)$. The positive frequency dispersion relation is 
\begin{equation}\label{dispersion_relation_cone}
\o =  \sqrt{p^2 +k^2}
\end{equation}
where $p=\abs{{\bf p}}$ is the momentum in the Minkowski directions.  Note that while $\o$ doesn't have an explicit $n$-dependence, the quantized $k$'s do depend on $n$ non-trivially.  
 
To make the problem more tractable, we regulate the infinite spatial volume of the cone by truncating the space at an outer coordinate $r_\text{out}\gg  r_b$, and impose there a Dirichlet boundary condition on the field. At the appropriate place in the calculation we take the limit $r_\text{out}\to +\infty$ to recover the full spacetime. 

For the interior solutions we use Bessel functions and find that for $n\neq0$ the only ones that are normalizable are
\begin{align}\label{Cone_interior_soln}
R_{n\neq0,k}(r)&=J_{\n}\(k r\) ~~~~~( 0\leq r \leq  r_b)
\end{align}
while, for $n=0$ we are allowed (See Appendix \ref{solutions_cone})
\begin{align}
R_{n=0,k}(r)&=c_1 J_{\n}\(k r\) +c_2 N_{\n}\(k r\)~~~~~( 0\leq r \leq  r_b),
\end{align}
From here on we set $c_2=0$, however we find no physical argument as to why this must necessarily be true\textsuperscript{\footnotemark[8]}\footnotetext[8]{In fact, we must only ensure that the Hamiltonian is Hermitian and that the modes are normalizable; there is nothing inherently wrong with singular modes unless, of course, physical quantities diverge. The ambiguity in the value of $c_2$ represents an ignorance of what lies at the cone's tip; presumably it would be determined uniquely if the structural details of the tip were resolved.}
. For the exterior solutions we use Hankel functions for later convenience, and find (for all $n$)
\begin{align}\label{Cone_exterior_soln}
R_{n,k}(r)&=H^{(2)}_{\n}\(k r_\text{out}\) H^{(1)}_{\n}\(k r\) -H^{(1)}_{\n}\(k r_\text{out}\) H^{(2)}_{\n}\(k r\)~~~~~(  r_b\leq r \leq r_\text{out}),
\end{align}
where we have already imposed the Dirichlet condition at $r_\text{out}$. The spectrum of $k$ are determined by the boundary conditions at $ r_b$:
\begin{equation}
0 = 
\begin{cases}
 R_{n,k}( r_b)& \mbox{(Dirichlet)}\,,\\
   R_{n,k}'( r_b) &\mbox{(Neumann)}\,.
\end{cases}
\end{equation}

\sssec{Procedure for Calculation of Casimir Energy on the Cone}
~\\
\emph{Contour Integral Representation of Sums}\label{gen_contour_rep_cone}\\

Again, we have no explicit expression for the $\{k\}$, and so we calculate \eqref{zeta_fn_reg_E_0} in part using the contour integral representation
\begin{equation}\label{contour_rep_cone}
\sum_{\{k\}}k^{1-2s + m}=-\f{\cos{\p \(\f{m}{2}-s\)}}{\p}\int_{0}^{\infty} dk ~k^{1-2s + m}\f{\d}{\d k} \ln \Delta_n(i k)
\end{equation}
where the contour has already been deformed. Lastly, we must specify the mode generating functions, which are derived analogously to the case of the horn (see Appendix \ref{Cone_gen_fn}).  We find these functions for both sides of $ r_b$ to effectively combine to give the log of the total generating function

\begin{equation}
\label{logDeltaBoth_cone}
\ln{\Delta_n(i k)} = 
\begin{cases}
\ln{\[I_{\n}\(k  r_b\)K_{\n}\(k r_b\)\]}& \mbox{(Dirichlet)}\,,\\
\ln{\[I_{\n}'\(k  r_b\)K_{\n}'\(k r_b\)\]}&\mbox{(Neumann)}\,.
\end{cases}
\end{equation}
where any $ r_b$-independent terms have been omitted as they play no role in the Casimir effect.
\R
\R
\emph{General Analysis of $E_0(s)$}\\

We need to separate out the divergent parts from $E_0(s)$ so that they may be either analytically continued in $s$ or explicitly absorbed through some renormalization(s).  Because the summand is even in $n$, we will have contributions to the energy of the form
\begin{equation}\label{E_0_gen_cone_n_eq_0}
E^{(n=0)}_0\propto \int_{0}^{\infty} dk ~k^{1-2s+m}\f{\d}{\d k} \ln{\Delta_0(i k)}
\end{equation}
and
\begin{equation}\label{E_0_gen_cone_n_neq_0}
E^{(n\neq0)}_0\propto \sum_{n=1}^\infty \int_{0}^{\infty} dk ~k^{1-2s+m}\f{\d}{\d k} \ln{\Delta_n(i k)}
\end{equation}

Given \eqref{logDeltaBoth_cone}, an analytic continuation of these expressions is impossible in its current form. We want to decompose $E_0(s)$ into a sum of its asymptotic and finite part:
\begin{equation}
E_0(s) \equiv E_0^{\text{as}}(s)+E_0^{\text{\text{fin}}}
\end{equation}

For $n=0$ this simply means performing an expansion of $\ln{\Delta_0(i k)}$ in inverse powers of $k$.  Following \cite{Arfken}, we use the large $k$ expansion to define
\begin{align}
\widetilde{\ln} \[K_{0}\(k  r_b\)I_{0}\(k  r_b\)\] &\equiv  \ln{\[\f{1}{2k r_b}\]} +\sum_{j=1}^{m+1}\f{A_j}{\(k r_b\)^j}&&\text{(Dirichlet)}&\\
\widetilde{\ln}\[K_{0}'\(k  r_b\)I_{0}'\(k  r_b\)\] &\equiv  \ln{\[-\f{1}{2k r_b}\]} +\sum_{j=1}^{m+1}\f{B_j}{\(k r_b\)^j}&&\text{(Neumann)}&
\end{align}
where the first few coefficents are $A_2=1/8, A_4=13/64$, $B_2=-3/8, B_4=-27/64$ and $A_j=B_j=0$ for all odd $j$.
We thus capture all the divergences by taking 
\begin{equation}
E^{\text{as~} (n=0)}_0\propto \int_{0}^{\infty} dk ~k^{1-2s+m}\f{\d}{\d k} \widetilde{\ln} \[K_{0}\(k  r_b\)I_{0}\(k  r_b\)\]
\end{equation}
for the Dirichlet case, and similarly for Neumann.
For $n\neq0$, we employ a uniform asymptotic expansion (as done for the horn) by first making the variable change
\begin{equation}
y \equiv \f{k}{\n}
\end{equation}
%
therefore our sum/integral becomes
 \begin{equation}\label{E_0_gen_2_cone}
E^{(n\neq0)}_0\propto\sum_{n=1}^\infty \n^{1-2s+m}\int_{0}^{\infty} dy ~y^{1-2s+m}\f{\d}{\d y} \ln{\Delta_n(i\n y)}
\end{equation}
From the uniform asymptotic behavior of the two modified Bessel functions (see e.g. \cite{abramowitz+stegun}) we define the asymptotic part of the log of the generating function to be 
\begin{align}
\widetilde{\ln}\[K_\n\(\n y r_b\)I_\n\(\n y r_b\)\] &\equiv  \ln{\[\f{1}{2\n}\f{1}{\sqrt{y^2 r_b^2+1}}\]} +\sum_{j=1}^{m+2} \f{f_{j}(y  r_b)}{\n^{j}}~~~~~~\text{(Dirichlet)}\label{log_KI_expansion_cone}\\
\widetilde{\ln} \[K_\n'\(\n y r_b\)I_\n'\(\n y r_b\)\] &\equiv  \ln{\[-\f{1}{2\n}\f{\sqrt{y^2 r_b^2+1}}{y^2 r_b^2}\]} +\sum_{j=1}^{m+2} \f{h_{j}(y  r_b)}{\n^{j}}~~~~~~\text{(Neumann),}\label{log_KpIp_expansion_cone}
\end{align}
where the $f_{j}(z)$ and $h_{j}(z)$ are given in Appendix \ref{Bessel_asymptotics}.  We therefore capture all the divergences of \eqref{E_0_gen_2_cone} by taking
\begin{align}\label{E_0_as_cone}
E^{\text{as~} (n\neq0)}_0 \propto  \sum_{n=1}^\infty & \n^{1-2s+m}\int_{0}^{\infty} dy ~y^{1-2s+m} \f{\d}{\d y} \widetilde{\ln}\[K_\n\(\n y r_b\)I_\n\(\n y r_b\)\].
\end{align}
for the Dirichlet case, and similarly for Neumann. 
In all cases, $E_0^{\text{fin}}$ is obtained from the difference of $E_0(s)$ and $E_0^{\text{as}}(s)$  To recap, we have in total
\begin{align}\label{total_E_cone}
E &= E^{~}_{\text{ten}} + E_\text{curv} + E_0^{\text{as}}(s)+E_0^{\text{fin}}  ~~~~~~&(D=2+1)\\
\r&=\r^{~}_{\text{ten}} + \r_\text{curv} + \r_0^{\text{as}}(s) + \r_0^{\text{fin}}  + \r_\text{0, SM}&(D=5+1)
\end{align}

\ssec{$D=2+1$}\label{section:2+1D_cone}


Here there are no Minkowski dimensions, so ${\bf p}=0$ and $m=0$. From \eqref{zeta_fn_reg_E_0} and \eqref{contour_rep_cone} we find 
\begin{equation}\label{E0_3D_cone}
E_0(s)=-\f{\m^{2s}}{2\p}\cos{\p s}\( \int_{0}^{\infty} dk ~ k^{1-2s}\f{\d}{\d k} \ln{\Delta_0(i k)} +2 \sum_{n=1}^\infty \int_{0}^{\infty} dk ~ k^{1-2s}\f{\d}{\d k} \ln{\Delta_n(i k)}\)
\end{equation}

\begin{figure}
  \begin{center}
    \includegraphics[scale=.5]{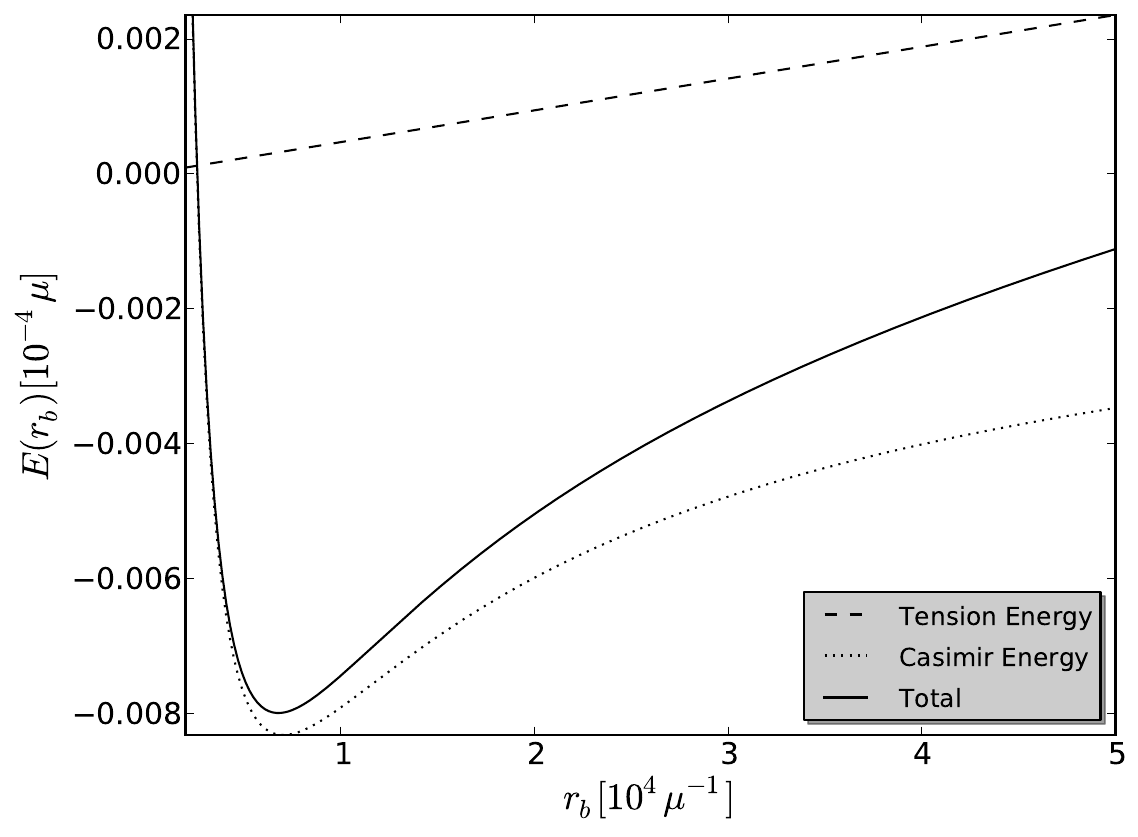}
  \end{center}
\caption{For $D=2+1$ with the Dirichlet boundary condition on $\phi$, the total energy in units ($\m$) where $h_\text{ren}=0$, as a function of the boundary position, $ r_b$. For illustration, we have taken $\de=1/4$ and the brane tension, $\s =10^{-12}~[\m^2]$.}
\label{3DConeDirichlet}
\end{figure}

\sssec{Dirichlet Condition}
\emph{The $n=0$ modes}:
In this case 
%
we isolate the divergent part as 
\begin{align}
E^{\text{as~} (n=0)}_0=-\f{\m^{2s}}{2\p}\cos{\p s} \int_{\L_\text{IR}}^{\infty} dk ~ k^{1-2s}\f{\d}{\d k} \widetilde{\ln} \[I_{0}\(k  r_b\)K_{0}\(k r_b\)\]
\end{align}
which vanishes after taking both $\L_\text{IR}\to0$ and $s \to 0$, the order being irrelevant. The finite piece is then found numerically (after integration by parts)
\begin{align}
E^{\text{fin~} (n=0)}_0&=+\f{1}{2\p} \int_{0}^{\infty} dk ~ \[\ln{\[I_{0}\(k  r_b\)K_{0}\(k r_b\)\]}-\widetilde{\ln} \[I_{0}\(k  r_b\)K_{0}\(k r_b\))\]\]\notag\\
&\approx - \f{1.4 \times 10^{-2}}{ r_b}
\end{align}
\R
\emph{The $n\neq0$ modes}:
Here we have (with $k=\n y$)
\begin{align}
E^{\text{as~} (n\neq0)}_0 &=-\f{\m^{2s}}{\p}\cos{\p s}  \sum_{n=1}^\infty  \n^{1-2s}\int_{0}^{\infty} dy ~y^{1-2s} \f{\d}{\d y} \widetilde{\ln}{\[I_{\n}\(\n y  r_b\)K_{\n}\(\n y  r_b\)\]}\\
&=\f{1}{24\(1-\de\)  r_b} - \f{\(1-\de\)}{128 r_b}\(3+\g + \f{1}{2s} + \ln{\[\m  r_b\(1-\de\)\]}\)
\end{align}
The finite piece is
\begin{align}
E^{\text{fin~} (n\neq0)}_0 &=+\f{1}{\p}  \sum_{n=1}^\infty  \int_{0}^{\infty} dk ~ \[\ln{\[I_{\n}\(k  r_b\)K_{\n}\(k r_b\)\]}-\widetilde{\ln}{\[I_{\n}\(k  r_b\)K_{\n}\(k  r_b\)\]}\]\\
&= \f{{\cal N}(\de)}{ r_b}
\end{align}
where ${\cal N}(\de)$ is a numerical coefficient; we find ${\cal N}(0) \approx 9.7 \times 10^{-4}$ (when the cone is opened fully) and that its magnitude decreases to zero\textsuperscript{\footnotemark[9]}\footnotetext[9]{This is consistent with the fact that in this limit $\widetilde{\ln}[\cdot]$ becomes a better approximation to ${\ln}[\cdot]$ since it is defined as an expansion in large $\n$.} monotonically as $\de \to 1$ (in the limit the cone closes in on itself). 
\R
\emph{Summary}:
A $1/s$ pole remains whose existence we verify using a heat kernel analysis in Appendix \ref{heat_appendix} and, based on how this term scales, we identify $h$ as the parameter that must be renormalized. Taking the total energy as the sum of the tension energy, the renormalized curvature, and the vacuum contributions we find
\begin{align}
E( r_b)
\approx 2\p \(1-\de\) r_b  \s  +\f{2\p\(1-\de\)h_\text{ren}}{ r_b} - \f{1.3 \times 10^{-2}}{ r_b} + \f{1}{24}\f{1}{\(1-\de\)  r_b} - \f{\(1-\de\)\ln{\m r_b}}{128 r_b} 
\end{align}
where we have also absorbed finite quantum corrections in $h_\text{ren}$. The scale $\mu$ remains in the argument of the logarithm, as it must for it be unitless. This scale, however, is arbitrary and completely degenerate with the measured value of $h_\text{ren}$, and may be chosen to be anything convenient. Then, without loss of generality, we define $\m$ as the scale at which $h_\text{ren}=0$. The logarithmic term dominates in both the $ r_b \to 0$ and $ r_b \to \infty$ limits and, because of its overall minus sign, the energy has a global minimum (see Figure \ref{3DConeDirichlet}).  Having $\s >0$ simply steepens the potential, however it isn't required for stability.

Finally, as a check on our results, we set $\s=\de=0$, to recover the scalar Casimir energy of a 1-sphere with a Dirichlet boundary condition (e.g. \cite{Bordag:2009zz,Leseduarte:1996ah})
\begin{equation}
\m^{-1}E_0(\m^{-1})\approx 6.8 \times 10^{-4}
\end{equation}

\sssec{Neumann Condition}

Here we proceed as in the previous section, using \eqref{logDeltaBoth_cone}.  We find the total energy to be
\begin{align}
E( r_b)
\approx 2\p \(1-\de\) r_b  \s  +\f{2\p\(1-\de\)h_\text{ren}}{ r_b}  - \f{0.25}{ r_b} -\f{1}{24}\f{1}{\(1-\de\)  r_b}  - \f{5\(1-\de\)}{128 r_b}\ln{ \m r_b}
\end{align}

\begin{figure}
  \begin{center}
    \includegraphics[scale=.5]{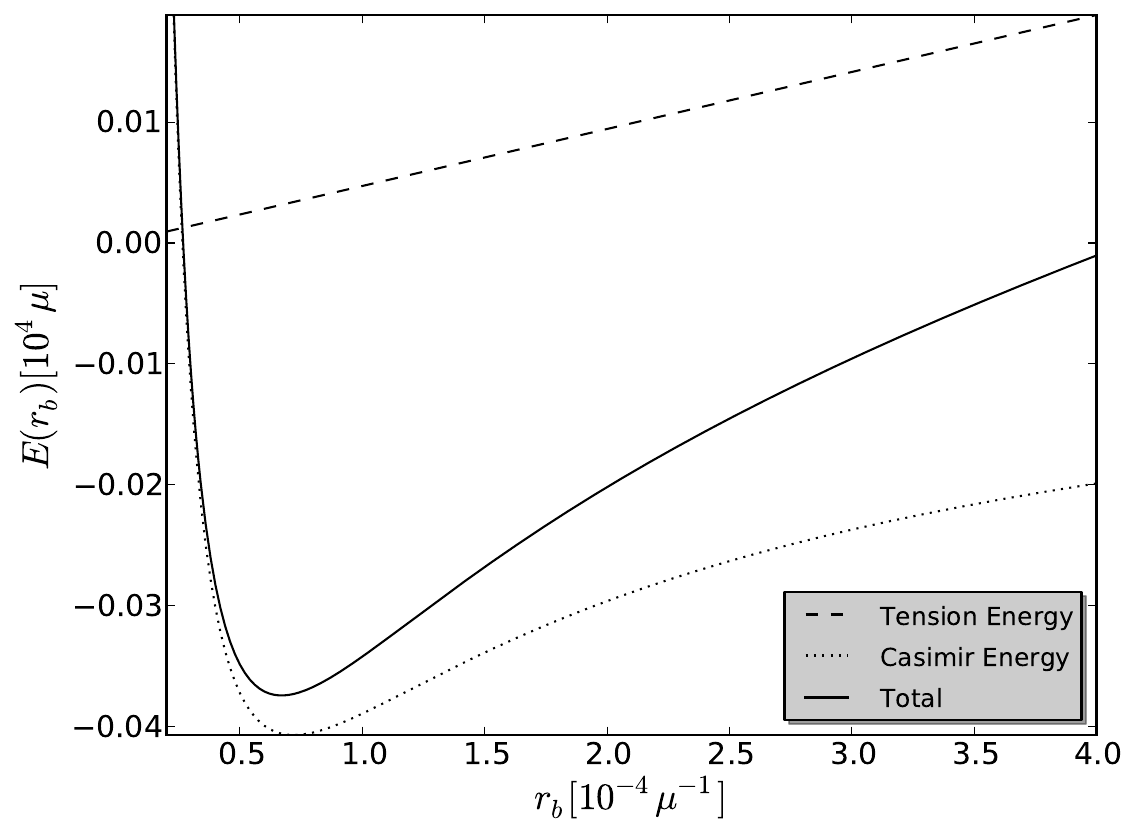}
  \end{center}
\caption{For $D=2+1$ with the Neumann boundary condition, the total energy in units ($\m$) where $h_\text{ren}=0$, as a function of the boundary position, $ r_b$. For illustration, we have taken $\de=1/4$ and the brane tension, $\s =10^{5}~[\m^2]$.}
\label{3DConeNeumann}
\end{figure}

As in the Dirichlet case, the log term dominates in both limits of $ r_b$, meaning there is again a global minimum\textsuperscript{\footnotemark[10]}\footnotetext[10]{One may notice a large difference in scales between the Dirichlet and Neumann case, namely that the minima lie at around $10^4 \m^{-1}$ and $10^{-4} \m^{-1}$, respectively. This is traced back to the logarithmic behavior of the potential, therefore the minimum depends exponentially on ${\cal O}(1)$ numbers that are either $>0$ or $<0$.} in the potential (see Figure \ref{3DConeNeumann}). As a check on our results, we set $\s=\de=0$, to recover the scalar Casimir energy of a 1-sphere with a Neumann boundary condition \cite{Bordag:2009zz,Leseduarte:1996ah}
\begin{equation}
\m^{-1}E_0(\m^{-1})\approx -0.18
\end{equation}

\ssec{$D=5+1$}\label{section:5+1D_cone}
Here, we treat the addition of three Minkowksi spatial dimensions ($m=3$) in the same manner as Section \ref{section:5+1D}, regulating the size of the Minkowksi spatial slice by  compactifying these dimensions using a torus of fundamental side length, $V_\text{M}^{1/3}$. Using the same logic and 
the result \eqref{contour_rep_cone} we find the effective 4D vacuum energy
\begin{align}
\r_0\(s\)&=\f{\m^{2s}}{16\p^{5/2}}\f{\Gamma(s-2)}{\Gamma(s-1/2)}\sin{\p s}\sum_{n} \int_0^\infty dk~   k^{4-2s}\f{\d}{\d k} \ln{\Delta_n(ik)}\notag\\
&=-\f{\m^{2s} }{64\p^2}\(1+s\(\f{3}{2}-\g -\psi(\sfrac{-1}{2})\)\)\times\notag\\
&~~~~~~~~~~~~~\(\int_0^\infty dk~   k^{4-2s}\f{\d}{\d k} \ln{\Delta_0(ik)} + 2\sum_{n=1}^\infty \int_0^\infty dk~   k^{4-2s}\f{\d}{\d k} \ln{\Delta_n(ik)}\)
\end{align}

%

\begin{figure}
  \begin{center}
    \includegraphics[scale=.5]{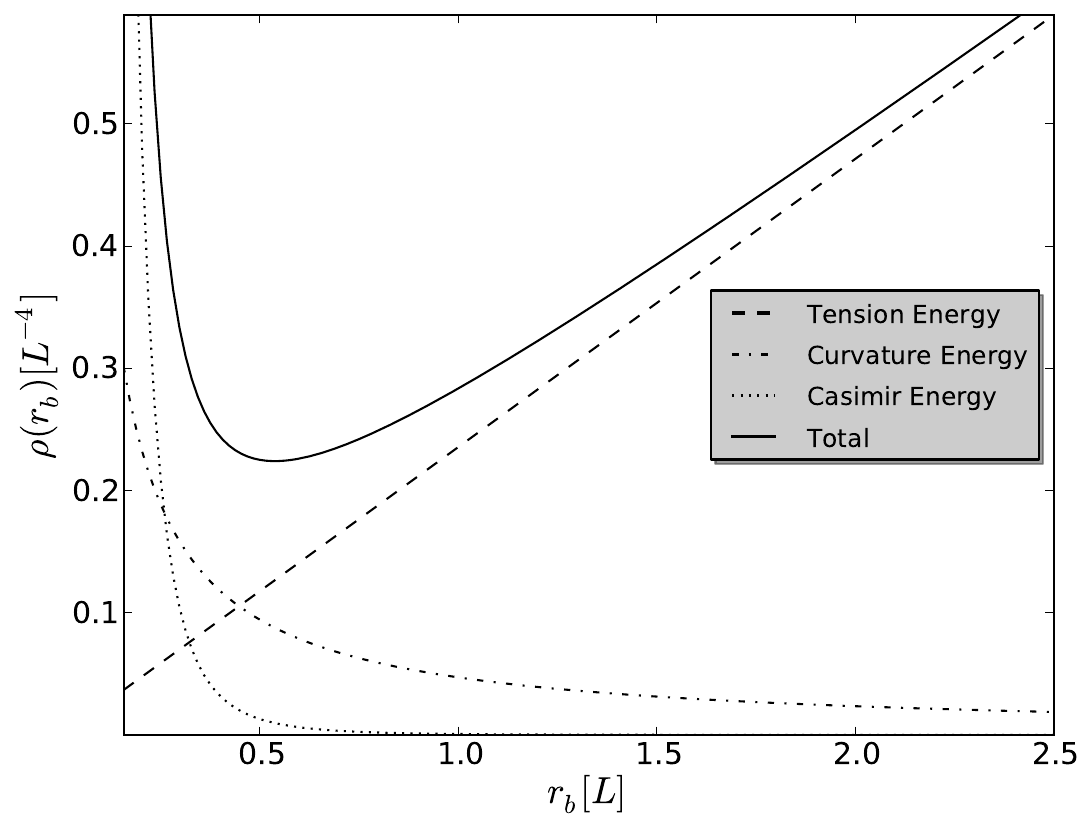}
  \end{center}
\caption{For the $D=5+1$ Dirichlet condition, the total 4D energy density vs. brane position using an arbitrary unit of length, $L$,  as a function of the boundary position, $ r_b$. For illustration, we have taken $\de=1/4$, $\s =0.05~[L^{-5}]$,  $h=0.01~[L^{-3}]$, $\k_{SM}=0$.}
\label{6DConeDirichlet}
\end{figure}

\sssec{Dirichlet Condition}
\emph{The $n=0$ modes}:
%
%
Here we define the analytic part as
\begin{align}
\r^{\text{as~} (n=0)}_0&=-\f{\m^{2s}}{64\p^2}\(1+s\(\f{3}{2}-\g -\psi(\sfrac{-1}{2})\)\) \int_{\L_\text{IR}}^{\infty} dk ~ k^{4-2s}\f{\d}{\d k}\widetilde{\ln} \[I_{0}\(k  r_b\)K_{0}\(k r_b\)\]\notag\\
&=\f{13}{2048\p^2  r_b^4}\(\f{1}{s}+2\ln{\m r_b} +\f{15}{26} - \g-\psi(\sfrac{-1}{2})\) 
\end{align}
where in the second line we have chosen\textsuperscript{\footnotemark[11]}\footnotetext[11]{In the $D=2+1$ cases we were able to simply analytically continue $\L_\text{IR}\to 0$. Here we must choose it to be finite (but arbitrary) and $\L_\text{IR}= r_b^{-1}$ is convenient for displaying our results. To be clear, the sum $\r^{(n=0)}_0=\r^{\text{as~} (n=0)}_0 + \r^{\text{fin~} (n=0)}_0$ is independent of $\L_\text{IR}$ in the $s\to 0$ limit.} $\L_\text{IR}= r_b^{-1}$.  The finite piece is then
\begin{align}
\r^{\text{fin~} (n=0)}_0&=-\f{1}{64\p^2}\( \int_{0}^{\L_\text{IR}} dk ~k^4 \f{\d}{\d k}\ln{\[I_{0}\(k  r_b\)K_{0}\(k r_b\)\]}\right.\notag\\
&~~~~~~~~~~~~~~+\left. \int_{\L_\text{IR}}^{\infty} dk ~k^4 \f{\d}{\d k}\[\ln{\[I_{0}\(k  r_b\)K_{0}\(k r_b\)\]}-\widetilde{\ln} \[I_{0}\(k  r_b\)K_{0}\(k r_b\)\]\]\)\notag\\
&\approx - 5.4\times 10^{-4}\times\f{1}{ r_b^4}
\end{align}
\R
\emph{The $n\neq0$ modes}:
Here, with $k=\n y $ we have
\begin{align}
\r^{\text{as~} (n\neq0)}_0 =&-\f{\m^{2s} }{32\p^2}\(1+s\(\f{3}{2}-\g -\psi(\sfrac{-1}{2})\)\)  \sum_{n=1}^\infty  \n^{4-2s}\int_{0}^{\infty} dy ~y^{4-2s} \f{\d}{\d y} \widetilde{\ln}{\[I_{\n}\(\n y  r_b\)K_{\n}\(\n y  r_b\)\]}\notag\\
&=\f{13 }{2048 \p^2 r_b^4}\(\g - \f{1}{s}-2\ln{\[2\p(1-\de) r_b \m\]}+\psi(\sfrac{-1}{2})\) + \f{199 }{8192\p^2  r_b^4}\notag\\
&~~~~~~~~~~~~~~~~ - \f{7\zeta'(-2) }{64\p^2 \(1-\de\)^2 r_b^4} + \f{\zeta'(-4) }{32\p^2 \(1-\de\)^4 r_b^4} 
\end{align}
Leaving us with a finite part given by
\begin{align}
\r^{\text{fin~} (n\neq0)}_0 &=+\f{1}{8\p^2}  \sum_{n=1}^\infty  \int_{0}^{\infty} dk ~k^3 \[\ln{\[I_{\n}\(k  r_b\)K_{\n}\(k  r_b\)\]} - \widetilde{\ln}{\[I_{\n}\(k  r_b\)K_{\n}\(k  r_b\)\]}\]\\
&\approx {\cal N}(\de)\times\f{1}{ r_b^4}
\end{align}
where ${\cal N}(\de)$ is a numerical coefficient. It is about $-2.8 \times 10^{-4}$ for $\de=0$ and its magnitude decreases to zero monotonically as $\de$ increases towards $1$. 
\R\\
\emph{Summary}:
Taking the total energy as the sum of the tension, renormalized curvature, and vacuum energy contributions, we have

\begin{align}
\r(r_b)
\approx   &~ 2\p \(1-\de\) r_b  \s + \f{2\p \(1-\de\)h_\text{ren}}{ r_b} +\f{229}{8192 \p^2 r_b^4}-\f{13\ln{[2\p(1-\de)]}}{1024 \p^2 r_b^4}\\  
&~~~~~ - \f{7\zeta'(-2)}{64\p^2 \(1-\de\)^2 r_b^4} +\f{1}{16\p^4 \(1-\de\)^4 r_b^4}\(\f{1}{2}\p^2 \zeta'(-4) + \k_{SM}\)\notag
\end{align}
where $1/2\p^2 \zeta'(-4) \approx 0.04$.  Here the $1/s$ poles have canceled when the $n=0$ and $n\neq0$ modes were added together. As $ r_b\to 0$ the energy is unbounded from above for all $\de$,
 neglecting $\k_{SM}$. Therefore, if a positive brane tension exists the energy has a stable minimum.  Since the cone has no intrinsic length scale we choose use an arbitrary scale, $L$, as a reference with which to plot our results in Fig \ref{6DConeDirichlet}. Interestingly, even in the absence of the Casimir effects of the bulk scalar, a stable minimum is achieved if we take both $h$ and $\s$ to be positive, a physically reasonable assumption.

\begin{figure}
  \begin{center}
    \includegraphics[scale=.5]{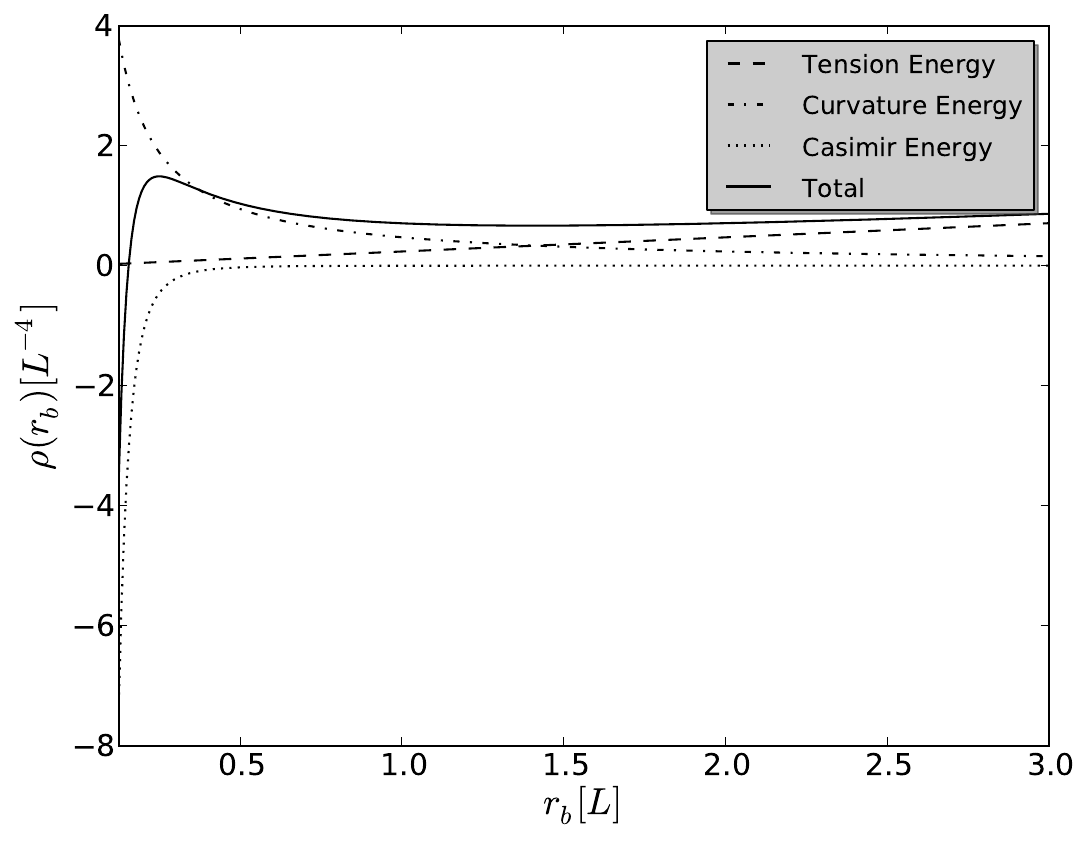}
  \end{center}
\caption{For the $D=5+1$ Neumann condition, the total 4D energy density using an arbitrary unit of length, $L$, as a function of the boundary position, $ r_b$. For illustration, we have chosen $\de=1/4$, $\s =0.05~[L^{-5}]$, $h=0.1~[L^{-3}]$, $\k_{SM}=0$.}
\label{6DConeNeumann}
\end{figure}

\sssec{Neumann Condition }
\emph{Summary}: He we proceed as in the previous section, only using \eqref{logDeltaBoth_cone}. The total 4D energy density is found to be
\begin{align}
\r(r_b) \approx ~&   2\p \(1-\de\) r_b  \s + \f{2\p \(1-\de\)h}{ r_b}- \f{7309}{122880\p^2  r_b^4} +\f{27\ln{[2\p\(1-\de\)]}}{1024\p^2 r_b^4}\\
& + \f{13\zeta'(-2)}{64\p^2 \(1-\de\)^2 r_b^4} + \f{1}{16\p^4 \(1-\de\)^4 r_b^4}\(-\f{1}{2}\p^2 \zeta'(-4) + \k_{SM}\)\notag
\end{align}
Here the energy is unbounded from below as $ r_b\to0$ for all $\de$, neglecting $\k_{SM}$. A global minimum apparently doesn't  exist when $\kappa_{SM}$ is negligible, however because of the different scaling behavior of each contribution to the potential, a local minimum is possible for large enough $h$.  We have plotted this case in Fig \ref{6DConeNeumann}, again using a reference scale, $L$.

\section{Conclusions}\label{conclusions}

In this work we have generalized the analysis in \cite{Jacobs:2012ph}, demonstrating that a single brane embedded in a bulk spacetime can be localized and stabilized using various contributions to its effective potential, specifically exploiting the non-trivial geometry and/or topology of the bulk manifold. We have focused on the hyperbolic horn and Euclidean cone geometries as they have features that are generic to more general manifolds yet still allow our calculations to be tractable.  We have considered a brane wrapped around both manifolds, calculating the energy due to tension and extrinsic curvature (i.e. due to local brane geometry) and also due to the Casimir energy of a bulk scalar (a global effect).  Both types of contributions are generally regional in nature, most sensitive to a finite region within the spatial manifold. The spaces considered here might be taken seriously for model building, as an approximation to capture the \emph{regional} features of a more generic manifold, or perhaps simply as a toy example of how various forces can be used to stabilize branes embedded within non-trivial manifolds. 

We have considered both Dirichlet and Neumann boundary conditions for the bulk scalar at the location of the brane in both dimensions $D=2+1$ and $D=5+1$, i.e. without and with 3 Minkowski dimensions, respectively. Though we don't appear to live in $D=2+1$, those systems are interesting as they illustrate how the Casimir effect changes its behavior with the number of dimensions, perhaps most notably how the explicit presence of poles in the energy depend on whether $D$ is even or odd.  In Table \ref{table:results} we summarize our results on the type of minima found for each brane potential, assuming the geometric energies are positive ($\s, h >0$) and that the Casimir energy of Standard Model fields in $D=5+1$ are negligible ($\kappa_{SM} \ll 1$).

\begin{table}[ht]
\centering  
\begin{tabular}{c c c c c} 
\hline\hline                        
Case &\multicolumn{2}{c}{$D=2+1$} & \multicolumn{2}{c}{$D=5+1$} \\[0.5ex]
~ & Dirichlet & Neumann & Dirichlet & Neumann \\
\hline                  
Horn & global & none & global & none \\ 
Cone & global & global & global & local \\[1ex]
\hline 
\end{tabular}
\caption{Summary of brane potentials, indicating whether a global, local or no minimum is possible, assuming $\kappa_{SM}$ is negligible.}
\label{table:results} 
\end{table}

Interestingly, for the cone we have demonstrated that a potential minimum is possible in the absence of Casimir energy; the competition between purely geometrical effects (tension and elastic energy) are sufficient to stabilize the brane. Physically, this is the result of tension energy being minimized by shrinking the volume of the brane, while the elastic energy is minimized by flattening it. On the cone, these two effects come in with markedly different dependences on the brane position, therefore an extremum is possible.  For the horn, however, these energies have the same brane position dependence and therefore the same cannot be said on that geometry.

Because the geometries we have chosen still retain a scaling symmetry, there are only two ``special" places on each manifold, i.e. $(-\infty, +\infty)$ or $(0, +\infty)$ on the horn and cone, respectively. As a result, all of the energy contributions behave monotonically with brane position and so are either minimized in the limit that the brane approaches one or the other. For this reason, we have found a necessary competition between at least two of them in order to achieve a stable minimum.

In fact, this is an especially important point in the case that we live on a 3-brane. In those scenarios the brane may be visualized (suppressing the 3 Minkowksi dimensions) as a point embedded in the bulk, having only trivial geometric properties that don't result in position-dependent forces. With (apparently) only the Casimir effect remaining, there is likely no global minimum\textsuperscript{\footnotemark[12]}\footnotetext[12]{However a local minimum may be possible by the competition of Casimir energies from several fields of varying mass and spin (see e.g. \cite{Greene:2007xu}).} for the brane potential on either the horn or the cone in this case and such a brane could only be driven to one of the extremes of the manifold. However this will not generically be the case; manifolds with sufficient symmetry violation will have many ``special" places -- places where one or many of the energies are extremized, especially Casimir.

Finally, we note that the discussion of regional effects of bulk manifolds is not limited to braneworld scenarios, alone.  Perhaps the most interesting (and generic) feature of non-trivial regional properties of extra-dimensional manifolds is the inhomogeneity of field modes of the bulk.  
\R
We would like to thank Ling-yi Huang, Harsh Mathur and Claudia de Rham for many useful discussions. This work is supported by grants from the U.S. Department of Energy and Department of Education.

\newpage

\bibliographystyle{JHEP_bibstyle}

\bibliography{Casimir_bib}

\newpage
\appendix

\sec{Details of Mode Solutions}\label{solutions}

\ssec{Hyperbolic Horn}\label{solutions_horn}

\sssec{$n\neq 0$}
From \eqref{horn_zeqn} we have
\begin{equation}
Z_{n,k}''(z) - Z_{n,k}'(z) + \(k^2 +\f{1}{4} - n^2e^{2z}\)Z_{n,k}(z)=0
\end{equation}
where we have defined $k$ by $\o\equiv\sqrt{p^2 +k^2 +\f{1}{4}}$. With the substitutions $\r = \abs{n} e^z$, $Z(\r)\equiv \r^{1/2} T(\r)$, and $k\equiv -i \n$ we obtain
\begin{equation}\label{mod_Bessel}
T_{\r\r}+ \f{1}{\r}T_{\r}- \(\f{\n^2}{\r^2} + 1 \)T  =0
\end{equation}
This familiar equation is the modified Bessel equation, whose two linearly independent solutions are the modified Bessel functions of the first and second kind, $I_\n(\r)$ and $K_\n(\r)$, respectively (see,  e.g.  \cite{Jackson}).  Returning to the original variables, we have
\begin{equation}
Z_{n,k}(z)=e^{z/2}\(c_1I_{i k}\(\abs{n}e^{z}\)+ c_2 K_{ik}\(\abs{n}e^z\)\),~~~~~~~~~ (z\geq z_b)
\end{equation}
In this region, the Klein-Gordon norm tells us
\begin{equation}
\int_{z_b}^{\infty} dz e^{-z}~Z_{\bf i}(z)^* Z_{\bf j}(z)=\de_{\bf i j}
\end{equation}
Asymptotically, $I_\m(x) \sim \abs{x}^{-1/2}e^x$ as $x\to \infty$, regardless of $\m$, making the solution non-normalizable unless we take $c_1=0$.  Furthermore, the modified Bessel functions $I_\m(x), K_\m(x)$ are monotonic for $\m\in \mathbb{R}$, meaning neither $K_{ik}\(\abs{n}e^z\)$ nor its derivative vanish for finite $z$. As the $k \leftrightarrow -k$ solutions are linearly related, this indicates that only real, $k>0$ are permissible.

In fact, $I_\n(\r)$ and $I_{-\n}(\r)$ also form a set of linearly independent solutions to \eqref{mod_Bessel}, so long as both $\n \notin \mathbb{Z}$ and $i \n \notin \mathbb{Z}$ \textsuperscript{\footnotemark[13]}\footnotetext[13]{One may worry about the possibility of $k \in \mathbb{Z}$, however those cases are contrived; the set of model parameters for which this is true form a set of measure zero.}.   For later convenience, we use
\begin{equation}
Z_{n,k}(z)=e^{z/2}\(c_1I_{ik}\(\abs{n}e^{z}\)+ c_2 I_{-ik}\(\abs{n}e^z\)\), ~~~~~~~~~~(z_L \leq z \leq z_b)
\end{equation}
where the two boundary conditions at $z_L$ and $z_b$ plus the normalization condition completely determine $c_1, c_2$ and the allowable $k$.

For the region $z_L \leq z \leq z_b$  the conditions necessary to have a zero at $z_L$ tell us (up to a normalization factor)
\begin{equation}
Z_{n,k}(z) = e^{z/2}\(I_{-i k}(\abs{n} e^{z_L}) I_{ik}(\abs{n} e^{z}) -I_{ik}(\abs{n} e^{z_L}) I_{-ik}(\abs{n} e^{z})\)
\end{equation}
In the case of Dirichlet boundary condition, the zero at $z_b$ then implies the relation
\begin{equation}
\f{I_{ik}(\abs{n} e^{z_b})}{I_{-ik}(\abs{n} e^{z_b})}=\f{I_{ik}(\abs{n} e^{z_L})}{I_{-ik}(\abs{n} e^{z_L})}.
\end{equation}
and there is a similar relation for the Neumann case. When $ik \in \mathbb{R}$, the monotonicity of the modified Bessel function implies this equation may only be satisfied in the trivial case, $z_L=z_b$. All things considered, this means only real, $k>0$ are allowed. In summary,
\begin{align}
Z(z)&=e^{z/2}\[I_{-i k}\(\abs{n} e^{z_L}\) I_{ik}\(\abs{n} e^{z}\) -I_{ik}\(\abs{n} e^{z_L}\) I_{-ik}\(\abs{n} e^{z}\)\]~&(z_L \leq z \leq z_b)\\
&=e^{z/2} K_{ik}\(\abs{n} e^{z}\) &(z_b \leq z)\notag
\end{align}

\sssec{$n=0$}

From \eqref{horn_zeqn} we have
\begin{equation}
Z''(z) -  Z'(z) + \(k^2+\f{1}{4}\)Z(z)=0
\end{equation}
The solutions are
\begin{equation}
Z(z)= e^{z/2}\(c_1e^{ik z}+ c_2 e^{-ik z}\) 
\end{equation}
To obviate having to delta function-normalize the modes in the region $z\geq z_b$, we truncate the space at some $z_{R} > z_b$, taking the limit $z_{R}\to \infty$ at the end of the calculation.  This quantizes the modes in this region as well. After imposing zero at $z_L$ and $z_R$ the solutions are 
\begin{align}\label{Z_soln_1}
Z_{0,k}&=e^{z/2}\sin{k(z-z_L)}~&(z_L \leq z \leq z_b)\\
&=e^{z/2} \sin{k (z-z_R)} &(z_{b} \leq z \leq z_{R})
\end{align}
Requiring either a Dirichlet or Neumann boundary condition in each region implies real, $k>0$.

\ssec{Euclidean Cone}\label{solutions_cone}
Consider the interior solutions
\begin{equation}
R_{n,k}(r)=c_1 J_{\n}\(k r\) +c_2 N_{\n}\(k r\)~~~~~( 0\leq r \leq  r_b),
\end{equation}
for arbitrary $n$.  Normalizability would tell us
\begin{equation}
\int_{0}^{ r_b} dr~\!r~\!R_{n,k}(r)^2=1
\end{equation}
If $c_2\neq0$, the most dangerous part of this integral near $r=0$ is
\begin{align}
\int_{0} dr~\!r~\!N_{\n}\(k r\)^2 &\propto \int_{0} dr~\!r~\!r^{-2\n}, &(\n\neq0)\\
&\propto \int_{0} dr~\!r~\!\(\ln{p}\)^2, &(\n=0)
\end{align}

The first integral diverges because $\n\geq1$, while the second one converges.  Therefore, on the basis of normalizibility, $c_2\neq0$ is possible only for $\n=0$ ($n=0$).
\sec{Calculating $E_0(s)$ via Contour Integrals}\label{contour_details}
\ssec{General Procedure}

We will ensure $\Delta_n(k)$ to be a meromorphic function, having only simple zeros at the location of the $k$'s in the spectrum.  Near the $i$th zero, the derivative of its logarithm produces the pole $(k-k_i)^{-1}$, plus other finite terms so that, by the Cauchy residue theorem, we have the relation
\begin{equation}
\sum_{\{k\}} \(k^2 + \L^2\)^{\f{m+1}{2}-s}=\f{1}{2\p i}\oint_\g dk \(k^2 + \L^2\)^{\f{m+1}{2}-s}\f{\d}{\d k} \ln \Delta_n(k)  
\end{equation}
The beauty of this technique is that only the eigenfunctions need to be understood analytically.

\begin{figure}
  \begin{center}
    \includegraphics[scale=1]{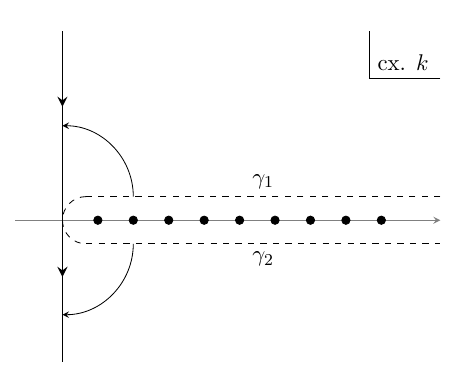}
      \end{center}
\caption{Deformed contour.}
\label{fig3}
\end{figure}

Assuming no other poles/zeros in the Re ${k}>0$ plane exist, the contour may be deformed (see Figure \ref{fig3}) by rotating counter-clockwise ($k\to e^{i\p/2}k$) the upper contour, $\g_1$, while the lower contour, $\g_2$, is rotated clockwise ($k\to e^{-i\p/2}k$). Thus the contour goes along the imaginary axis from $i\infty$ to $-i \infty$, discarding the rest of the boundary at $\abs{k}\to \infty$.  (This is legitimate so long as $s$ is large enough; at the end of the calculation we analytically continue to $s=0$). Care must be taken here as the function $\(k^2 + \L^2\)^{\f{m+1}{2}-s}$  is multivalued and has a branch point at $k=\pm \L$.
Explicitly, one finds
\begin{align} 
\f{1}{2\p i}&\oint_\g dk \(k^2 + \L^2\)^{\f{m+1}{2}-s}\f{\d}{\d k} \ln \Delta_n(k)\notag\\
 &= \f{1}{2\p i}\(\int_{i \infty}^{0} dk \(k^2 + \L^2\)^{\f{m+1}{2}-s}\f{\d}{\d k} \ln \Delta_n(k) + \int_{0}^{-i \infty} dk \(k^2 + \L^2\)^{\f{m+1}{2}-s}\f{\d}{\d k} \ln \Delta_n(k)\)\notag\\
&=\f{1}{2\p i}\int_0^{\infty} dk \[\(e^{-i \p}k^2 + \L^2\)^{\f{m+1}{2}-s} - \(e^{i \p}k^2 + \L^2\)^{\f{m+1}{2}-s} \]\f{\d}{\d k} \ln \Delta_n(ik)\notag\\
&=-\f{\cos{\p \(\f{m}{2}-s\)}}{\p}\int_{\L}^{\infty} dk \(k^2 - \L^2\)^{\f{m+1}{2}-s}\f{\d}{\d k} \ln \Delta_n(ik)  
\end{align}
where we have assumed that the chosen generating functions satisfy 
\begin{equation}
\f{\d}{\d k} \ln \Delta_n(ik)=\f{\d}{\d k} \ln \Delta_n(-ik)
\end{equation}

Below we shall give a brief derivation of the mode-generating functions for both the horn and cone geometries with the Dirichlet boundary condition. The extension to the Neumann condition is straightforward.

\ssec{Horn Generating Functions (Dirichlet, $n\neq0$)}

Here we need to calculate
\begin{equation}
\f{1}{2\p i}\oint_\g dk \(k^2 + \f{1}{4}\)^{\f{m+1}{2}-s}\f{\d}{\d k} \ln \Delta_n(k)
\end{equation}
Considering \eqref{horn_soln_nneq0_leftright} we choose the generating functions

\begin{align}
\Delta_n(k)&=\f{1}{k}\(I_{ik}\(\abs{n} e^{z_L}\) I_{-ik}\(\abs{n} e^{z_b}\)-I_{-i k}\(\abs{n} e^{z_L}\) I_{ik}\(\abs{n} e^{z_b}\)\),~~~~(z_L \leq z \leq z_b) \label{left_gen_fn_horn}\\
\Delta_n(k)&=K_{ik}\(\abs{n} e^{z_b}\),~~~~(z_b \leq z)\label{right_gen_fn}
\end{align}

It will be of no consequence to the actual calculation\textsuperscript{\footnotemark[14]}\footnotetext[14]{Since a logarithm is taken of the generating function, if we multiply it by any analytic function of $k$ this only contributes an additive (though possibly infinite) constant to $E_0$,  however this has no bearing on the Casimir force as it is $z_b$-independent.} but technically we had to explicitly include a factor of $k^{-1}$ in \eqref{left_gen_fn_horn} to avoid counting the zero mode in the spectrum (at small $k$ the term in parenthesis is proportional to $k$). As \eqref{left_gen_fn_horn} stands it is not useful, because we must still perform the limit $z_L\to -\infty$. Because of the asymptotic behavior $I_{\m}(x)\sim (x/2)^\m$ as $x\to0$, in the limit of $z_L \to - \infty$ the surviving $z_b$-dependent terms of \eqref{left_gen_fn_horn} are
\begin{align}
 I_{-ik}\(\abs{n} e^{z_b}\), ~~~~~\text{ on } \g_1~ (\text{Re}~k>0)\\
 I_{ik}\(\abs{n} e^{z_b}\),~~~~~\text{ on } \g_2~ (\text{Re}~k<0)
\end{align}

When the contour is deformed, these two terms are equal and proportional to $I_{k}\(\abs{n} e^{z_b}\)$.  Thus we effectively have
\begin{align}
\Delta_n(ik)&=I_{k}\(\abs{n} e^{z_b}\),~~~~(z_L \leq z \leq z_b)\\
\Delta_n(ik)&=K_{k}\(\abs{n} e^{z_b}\),~~~~(z_b \leq z)
\end{align}

Since the energy of each region is simply be added (as the energies of two volumes sitting adjacent to each other), the two generating functions combine to give
\begin{align}
\f{1}{2\p i}\oint_\g dk &\(k^2 + \f{1}{4}\)^{\f{m+1}{2}-s}\f{\d}{\d k} \ln \Delta_n(k)\notag\\
&=-\f{\cos{\p \(\f{m}{2}-s\)}}{\p} \int_{1/2}^{\infty} dk \(k^2 - \f{1}{4}\)^{\f{\(m+1\)}{2}-s}\f{\d}{\d k} \ln{\[I_{k}\(\abs{n} e^{z_b}\)K_{k}\(\abs{n} e^{z_b}\)\]}
\end{align}
\ssec{Horn Generating Functions (Dirichlet, $n=0$)}\label{Horn_gen_neq0}
Here we will show that the sum over the $k$-spectrum is $z_b$-independent, and can therefore be ignored in the calculation of the Casimir energy. Before rotating the contour, we have an integral of the form
\begin{equation}
\f{1}{2\p i}\oint_\g dk \(k^2 + \f{1}{4}\)^{\f{\(m+1\)}{2}-s}\f{\d}{\d k} \ln \Delta_0(k)
\end{equation}

We break up the regions and, considering \eqref{Z_soln_1}, choose the generating functions to be
\begin{align}
\Delta_0(k)&=\f{1}{k}\(e^{ik(z_b-z_L)}-e^{-ik(z_b-z_L)}\)&(z_L \leq z \leq z_b)\\
\Delta_0(k)&=\f{1}{k}\(e^{ik(z_R-z_b)}-e^{-ik(z_R-z_b)}\)&(z_b \leq z \leq z_R)\notag
\end{align}

First we consider the $z_L \leq z \leq z_b$ region, and take the limit $z_L \to -\infty$. On $\g_1$ (Im $k>0$), $\Delta_0(k) \to e^{-ik(z_b-z_L)}k^{-1}$. On $\g_2$ (Im $k<0$), $\Delta_0(k) \to e^{ik(z_b-z_L)}k^{-1}$.  The same procedure may be performed for the $z_b \leq z \leq z_R$ region, taking $z_R \to \infty$. After the contour rotation, the effective generating functions for the left and right regions become $e^{k(z_b-z_L)}$ and $e^{k(z_R-z_b)}$, respectively. Thus, because of the logarithm, when the regions are combined the \emph{$z_b$-dependence vanishes and so these modes are irrelevant for the Casimir effect}, as claimed.

%

\ssec{Cone Generating Functions} \label{Cone_gen_fn}

Considering \eqref{Cone_interior_soln} and \eqref{Cone_exterior_soln}, we choose the generating functions
\begin{align}
\Delta_n(k)&=k^{-\n}J_{\n}\(k  r_b\) &~~~~~( 0\leq r \leq  r_b)\\
\Delta_n(k)&=\(H^{(2)}_{\n}\(k r_\text{out}\) H^{(1)}_{\n}\(k  r_b\) -H^{(1)}_{\n}\(k r_\text{out}\) H^{(2)}_{\n}\(k  r_b\)\)&~~~~~(  r_b\leq r \leq r_\text{out})\label{out_gen_fn_cone}
\end{align}

The factor of $k^{-\n}$ in the top line is so that the zero mode is not technically counted (at small $k$ the term in parenthesis is proportional to $k^\n$). As it stands \eqref{out_gen_fn_cone} is not useful because we must still perform the limit $r_\text{out}\to \infty$. Because of the asymptotic behavior 

\begin{equation}
\lim_{x\to\infty}H_\n^{(1,2)}(x)\sim \sqrt{\f{2}{\p x}}e^{\pm i\(x -\n\p/2-\p/4\)},
\end{equation}
in the limit of $r_\text{out}\to \infty$ the surviving terms of \eqref{out_gen_fn_cone} are
\begin{align}
H^{(2)}_{\n}\(k r_\text{out}\) H^{(1)}_{\n}\(k  r_b\), ~~~~~\text{ on } \g_1~ (\text{Re}~k>0)\\
H^{(1)}_{\n}\(k r_\text{out}\) H^{(2)}_{\n}\(k  r_b\),~~~~~\text{ on } \g_2~ (\text{Re}~k<0)
\end{align}

When the contour is deformed, these two terms are both proportional to $I_{\n}\(k  r_b\)$, thus we effectively have
\begin{align}
\Delta_n(ik)&=I_{\n}\(k  r_b\)& &( 0\leq r \leq  r_b)\\
\Delta_n(ik)&=K_{\n}\(k  r_b\)& &(  r_b\leq r \leq r_\text{out})
\end{align}

Summing over the spectrum from each region we find
\begin{align}
\f{1}{2\p i}\oint_\g dk ~k^{1-2s + m}\f{\d}{\d k} \ln \Delta_n(k)=-\f{\cos{\p \(\f{m}{2}-s\)}}{\p} \int_{1/2}^{\infty} dk ~k^{1-2s + m}\f{\d}{\d k} \ln{\[I_{\n}\(k  r_b\) K_{\n}\(k  r_b\) \]}
\end{align}

\sec{Bessel Function Asymptotics}\label{Bessel_asymptotics}
Using the uniform asymptotic expansion of the modified Bessel functions (see e.g.\cite{abramowitz+stegun}), one may show for the Dirichlet boundary conditions
\begin{align}
\ln \[I_{\n}\(\n z\)K_{\n}\(\n z\)\] \sim  &\ln{\[\f{1}{2 \n}\f{1}{\sqrt{z^2+1}}\]} +\sum_{j=1}^\infty \f{f_{j}(z)}{\n^{j}}
\end{align}
where the first few non-zero $f_j(z)$ are
\begin{align}
f_2(z)&= \f{(-4+z^2)z^2}{8(1+z^2)^3}\\
f_4(z)&= \f{(-32+288z^2-232z^4+13z^6)z^2}{64(1+z^2)^6}\\
f_6(z)&= \f{(-48+2580z^2-14884z^4+17493z^6 - 4242z^8 + 103z^{10})z^2}{96(1+z^2)^9}
\end{align}
For the Neumann condition on the horn one finds
\begin{align}
&\ln{\[\f{1}{4}I_{\n}\(\n z\)K_{\n}\(\n z\) + \f{1}{2}\n z\(I_{\n}\(\n z\)K_{\n}'\(\n z\)+I_{\n}'\(\n z\)K_{\n}\(\n z\)\) + \(\n z\)^2 I_{\n}'\(\n z\)K_{\n}'\(\n z\)\]}\notag\\
&~~~~~~~~~~~~~~~~~~~~~~~~~ \sim  \ln{\[-\f{\n}{2}\sqrt{z^2+1}\]} +\sum_{j=1}^\infty \f{g_{j}(z)}{\n^{j}}
\end{align}
where the first few non-zero $g_j(z)$ are
\begin{align}
g_2(z)&= \f{-2+4z^2- z^4}{8(1+z^2)^3}\\
g_4(z)&= \f{-2 + 80z^2 -372z^4 +240z^6 -13 z^8}{64(1+z^2)^6}\\
g_6(z)&= \f{-1+276z^2 -8031z^4 +36182 z^6 -37659z^8 +8628z^{10} - 209 z^{12}}{192(1+z^2)^9}
\end{align}
Finally, for the Neumann condition on the cone, one may show
\begin{align}
\ln \[K_{\n}'\(\n z\)I_{\n}'\(\n z\)\] \sim  &\ln{\[-\f{1}{2 \n}\f{\sqrt{z^2+1}}{z^2}\]} +\sum_{j=1}^\infty \f{h_{j}(z)}{\n^{j}}
\end{align}
where the first few non-zero $h_j(z)$ are
\begin{align}
h_2(z)&= \f{(4-3z^2)z^2}{8(1+z^2)^3}\\
h_4(z)&= \f{(32 - 320z^2 + 328z^4 - 27z^6)z^2}{64(1+z^2)^6}\\
h_6(z)&= \f{(48 - 2652z^2 + 16180z^4 -20799 z^6 + 5652z^8 -162z^{10})z^2}{96(1+z^2)^9}
\end{align}

\sec{Details of $2+1$ Horn Calculation}\label{2+1_Horn_details}
\ssec{$E_0^{\text{div}}(s)$}
%
\emph{Renormalization of Brane Tension}:
%
We see in equation \eqref{E0rem_3D} that a pole remains\textsuperscript{\footnotemark[15]}\footnotetext[15]{See also Appendix \ref{heat_appendix} for an independent check using the heat kernel technique.}, specifically one that is $z_b$-dependent:
\begin{equation}
-\f{9}{256}e^{-z_b}\times \f{1}{s}
\end{equation}
The Casimir force cannot be infinite, so clearly we must identify a physical parameter to absorb this divergence.  Notice that this term is proportional to the volume of the brane
\begin{equation}
V_{\text{brane}} \propto \int_0^{2\p} d\th \sqrt{\abs{\g}} = 2\p e^{-z_b}=2\p e^{-z_b}
\end{equation}
This suggests that the brane tension is a renormalizable quantity that will suffice.  Let us consider the energy due to a bare brane tension, $\s$:
\begin{equation}
E_{\text{ten}}=\int_0^{2\p} d\th \sqrt{\abs{\g}} \s = 2\p e^{-zb}\times \s
\end{equation}
For convenience we define the renormalized brane tension using not only terms from $E_0^\text{div}$, but also the finite contributions to $E_0$ which are proportional to the brane volume (in the limit $z_b \to -\infty$):
\begin{equation}
E_\text{ten} \equiv 2 \p e^{-z_b} \s_\text{ren} \equiv 2 \p e^{-z_b}\( \s -\f{1}{2\p}\f{9}{128}\(-\f{13}{9} + \g + \f{1}{2s}  + \text{finite} \) \)
\end{equation}

With this definition,
\begin{equation}%
\lim_{z_b\to -\infty }E(z_b) \sim E_\text{ten}= 2\p e^{-zb}\times \s_\text{ren}\,,
\end{equation}
which is now finite and independent of $\m$.

\ssec{$E_0^{\text{\text{rem}}}$}
From \eqref{E0_rem_horn} we have
\begin{equation}\label{E0rem_3D_Horn}
E_0^{\text{\text{rem}}}= \sum_{n=1}^\infty\[-\f{1+16x_b^2 + 49 x_b^4 + 64 x_b^6}{4\(1+4 x_b^2\)^{5/2}} + \( \f{1}{2} x_b+ \f{9}{128}x_b^{-1}\)\]
\end{equation}
and remind the reader that $x_b=\abs{n}e^{z_b}$.\\
\\
{\bf Limit $z_b \to \infty$:} Since $x_b \gg 1$ for all $n$,  we can expand in $x_b^{-1}$ so that
\begin{align}
\lim_{z_b\to \infty}E_0^{\text{\text{rem}}}&\sim -\sum_{n=1}^\infty \f{23}{1024}x_b^{-3}
= -\f{23}{1024}\zeta(3)e^{-3z_b}
\end{align}
{\bf Limit $z_b \to -\infty$:} We can approximate the sum as an integral by using the Euler-Maclaurin formula which, to lowest order, says (see e.g. \cite{Press:2007:NRE:1403886})
\begin{align}\label{Euler-Mac}
\sum_{n=1}^\infty f(A n)&\approx \f{1}{A}\int_A^\infty dx f(x) + \f{1}{2}\(f\(A\)+f\(\infty\)\)
\end{align}
We identify $A\leftrightarrow e^{z_b}$ so that $f(x_b)$ is the summand of \eqref{E0rem_3D_Horn}, thus
\begin{align}
\lim_{z_b\to -\infty}E_0^{\text{\text{rem}}}&\sim  e^{-z_b}\int_{e^{z_b}}^\infty dx f(x) + \f{1}{2}f(e^{z_b})\notag\\
&\sim -\f{7 + 18\ln{[4e^{z_b}]}}{256}e^{-z_b} + \f{1}{8} + {\cal O}(e^{2z_b})\notag\\
\end{align}


\ssec{$E_0^{\text{fin}}$}
From \eqref{E0_fin_horn} we have
 \begin{equation}
E_0^{\text{fin}}=\f{1}{\p} \sum_{n=1}^\infty \int_{1/2}^{\infty} dk ~\f{k}{\(k^2 - \f{1}{4}\)^{1/2}} D_k(x_b)
\end{equation}
where
\begin{equation}
D_k(x_b)=\(\ln \[K_{k}\(x_b\)I_{k}\(x_b\)\] - \widetilde{\ln} \[K_{k}\(x_b\)I_{k}\(x_b\)\]\)
\end{equation}
\\
{\bf Limit $z_b \to \infty$:} Again, it will always be true that $x_b\gg1$. Under this condition, the uniform asymptotic expansions for the modified Bessel functions are a good approximation because beyond the leading logarithmic term they are expansions in inverse powers of $x_b$. Therefore by going to the next highest order in $x_b^{-1}$ we find (see Appendix \ref{Bessel_asymptotics}) 
\begin{equation}
D_k(x_b)\approx  \f{\(13x_b^6-232k^2x_b^4+288k^4x_b^2-32k^6\)x_b^2}{64(x_b^2+k^2)^6}
\end{equation}
This can be directly integrated and, after performing another expansion in $x_b^{-1}$
 \begin{align}
E_0^{\text{fin}}&\approx \sum_{n=1}^\infty \f{37\p}{16384}x_b^{-3} \notag
\\&=  -\f{37\p}{32768}\zeta(3)e^{-3z_b}
\end{align}
\\
{\bf Limit $z_b \to -\infty$:} Once again, we will use the lowest order Euler-Maclaurin expansion. In this limit one may show that
\begin{align}
\lim_{x\to 0}D_k(x) &\sim {\cal O}(x^2, x^{2k})\notag\\
\end{align}
depending on whether $k$ is less than or greater than $1$.  This implies that to good approximation
\begin{equation}\label{E0_fin_approx}
E_0^{\text{fin}} \approx \f{e^{-z_b}}{\p} \int_{1/2}^{\infty} dk ~\f{k}{\(k^2 - \f{1}{4}\)^{1/2}} \int_0^\infty dx ~D_k(x).
\end{equation}
In the above expression we have lowered the lower limit of integration to zero, the error being roughly
\begin{equation}\label{correction_3D}
e^{-z_b}\int_0^{e^{z_b}} dx ~D_k(x) \sim e^{-z_b}\int_0^{e^{z_b}} dx ~{\cal O}(x^2, x^{2k}) \sim {\cal O}(e^{2z_b}, e^{2kz_b})
\end{equation}
which is negligible as $z_b \to -\infty$. We have thus learned than in this limit $E_0^{\text{fin}} \propto e^{-z_b}$, so it remains to calculate this proportionality factor. Numerically computing the double integral in \eqref{E0_fin_approx}, we find
\begin{equation}
E^{\text{fin}} \approx 2.4\times10^{-3} \times e^{-z_b}
\end{equation}

\sec{Heat Kernel Analysis}\label{heat_appendix}

The zeta function is intimately related to the heat equation (see e.g. \cite{Elizalde:2007du}) as it is simply a Mellin transform of the heat operator. In this appendix we give a brief review of the heat kernel technique and apply it to confirm the poles encountered in $D=2+1$, providing a partial check on our calculations.

\ssec{Heat Kernel Analysis of a General $E_0(s)$}\label{heat_gen}
Following the nice review by Vassilevich \cite{Vassilevich:2003xt} and book by Fursaev and Vassilevich \cite{Fursaev:2011zz}, if we use a Mellin transform to write
\begin{equation}
\o_{\bf i}^{1-2s}=\int_0^\infty dt \f{t^{s-3/2}}{\Gamma\(s-\f{1}{2}\)} e^{-t \o_{\bf i}^{2}}
\end{equation}
then the zeta-function regularized Casimir energy \eqref{zeta_fn_reg_E_0} can be recast as

\begin{align}\label{E_0_w_K(t)}
E_0(s)&=\f{\m^{2s}}{2}\sum_{\bf i} \o_{\bf i}^{1-2s}\notag\\
&=\f{\m^{2s}}{2}      \int_0^\infty dt \f{t^{s-3/2}}{\Gamma\(s-\f{1}{2}\)}K(t)
\end{align}
where $K(t)$ is the heat kernel trace, given by\textsuperscript{\footnotemark[16]}\footnotetext[16]{Technically the summand should be $e^{-t\l_i}$, where the $\l_i$ are the eigenvalues of the generalized Laplace-Beltrami operator. Since we use a simple massless Klein-Gordon field, $\o_i^2=\l_i$.}
\begin{equation}\label{heat_trace}
K(t) = \sum_i e^{-t \o_i^2}=\sum_{p=0} t^{\f{p-n}{2}}a_p
\end{equation}
where $n$ is the number of spatial dimensions and the heat kernel coefficients, $a_p$, are determined by integrals over local geometric scalars and also depend on the field type and boundary conditions.  Considering \eqref{heat_trace}, in the limit $s\to0$ the divergent terms of \eqref{E_0_w_K(t)} are those in which $p\leq n+1$.  Analytically continuing $s\to0$ gives a finite answer for all terms except when $p=n+1$; this is an inescapable pole.  The divergence comes from the $t\to 0$ part of the integral, so we break the integral \eqref{E_0_w_K(t)} up into two regions, $[0,1]$ and $[1,\infty)$.  The later converges, while the former contains the pole:
\begin{align}\label{EPole}
 \f{\m^{2s}}{2}    \int_0^1 dt \f{t^{s-1}}{\Gamma\(s-\f{1}{2}\)}  a_{n+1} =&-\f{\m^{2s}}{4\sqrt{\p}}  a_{n+1}  \f{t^{s}}{s}\Big{|}^1_0 + {\cal O}(s^0)\notag\\
\sim&-\f{1}{4\sqrt{\p}}  a_{n+1} \f{1}{s} 
\end{align}

\ssec{Heat Kernel Technique Applied to the $D=2+1$ Geometries}
To verify the poles in the Casimir energy encountered for both the horn and cone geometries in $D=2+1$ ($n=2$) we require the coefficient $a_3$, for which we borrow the results from \cite{Vassilevich:2003xt}, suited for our problem and conventions.   To apply them, we identify the variables from that work,  ($f, E, S, \Pi_+, \Pi_-$) as $f=1$, $E=0$, and $S=0$, with $\Pi_+$ and $\Pi_-$ depending on the choice of boundary condition. With these results in hand, we have
\begin{equation}
a_3=\f{1}{768 \sqrt{\p}}\int_{\d {\cal M}}\!\! d^{n-1}x \sqrt{\abs{\g}} \(16\chi R - 8 R_{a n a n} +\(13\Pi_+ - 7\Pi_-\) K^2 +\(2\Pi_+ + 10\Pi_-\) K_{ab}K_{ab}\)
\end{equation}
where $\g_{\m\n}$ is the induced metric on the boundary, $\chi=\Pi_+ - \Pi_-$, the Riemann tensor is defined\textsuperscript{\footnotemark[17]}\footnotetext[17]{There is a difference of a minus sign between the Riemann tensor as defined in \cite{Vassilevich:2003xt} and here.} by $\[\del_\m,\del_\n\]A^\a \equiv R^\a_{~\l\m\n}A^\l$ , and $K_{ab}$ is the extrinsic curvature tensor of the brane.  Latin (or hatted) indices designate orthonormal coordinates, and repeated indices imply a summation.  As we have split up both of the geometries into two regions, we take care to calculate $a_3$ on each side of the boundary separately, however, the answers are identical in the cases we consider.

\sssec{Hyperbolic Horn}\label{heat_horn}
Here the determinant of the induced metric on the brane is given by $\abs{\g}=e^{-2z_b}$, where we have set the horn curvature scale, $z_\star\equiv 1$. The relevant geometric objects are
\begin{align}
R&=-2\\
R_{\hat{\th} \hat{z} \hat{\th} \hat{z}}&=-1\\
K_{ab}&=\pm \de_{a\hat{\th}}\de_{b\hat{\th}}
\end{align}
where the sign of $K_{ab}$ is positive (negative) if the normal to the boundary, $n^\m$ points in the positive (negative) $z$ direction.
\R
\emph{Dirichlet Condition}\\

Here we identify $\Pi_+=0, \Pi_-=1$. For either region we find
\begin{equation}
a_3=\f{9}{128} \sqrt{\p} e^{-z_b}
\end{equation}
Considering \eqref{EPole}, we confirm that the entire pole part of the energy is
\begin{equation}\label{EP_horn_D}
E^{\text P}_0(s)= -\f{9}{256}  e^{-z_b} \f{1}{s}
\end{equation}\\\\\\\\
\emph{Neumann Condition}\\

Here $\Pi_+=1, \Pi_-=0$. For either region we find
\begin{equation}
a_3=-\f{3}{128} \sqrt{\p} e^{-z_b}
\end{equation}
implying
\begin{equation}\label{EP_horn_N}
E^{\text P}_0(s)= \f{3}{256} e^{-z_b} \f{1}{s}
\end{equation}

\sssec{Euclidean Cone}\label{heat_cone}
Here, the determinant of the induced metric on the brane is given by $\abs{\g}= r_b^2$. As the geometry is Euclidean, only the extrinsic curvature tensor is non-zero:
\begin{align}
K_{ab}&=\pm \f{1}{ r_b}\de_{a\th}\de_{b\th}
\end{align}
where the sign of $K_{ab}$ is positive (negative) if the normal to the boundary, $n^\m$ points in the positive (negative) $r$ direction.\\\\
\emph{Dirichlet Condition}\\

Here $\Pi_+=0, \Pi_-=1$. For either region we find
\begin{equation}
a_3=\f{\(1-\de\)\sqrt{\p}}{128 r_b}  
\end{equation}
Considering \eqref{EPole}, we confirm the entire pole part of the energy is
\begin{equation}\label{EP_cone_D}
E^{\text P}_0(s)= -\f{\(1-\de\)}{256 r_b}  \f{1}{s}
\end{equation}\\
\emph{Neumann Condition}\\

Here $\Pi_+=1, \Pi_-=0$. For either region we find
\begin{equation}
a_3=\f{5\(1-\de\)\sqrt{\p}}{128 r_b}  
\end{equation}
implying
\begin{equation}\label{EP_cone_N}
E^{\text P}_0(s)= -\f{5\(1-\de\)}{256 r_b}  \f{1}{s}
\end{equation}

\end{document}